\documentclass[12pt,preprint]{aastex}

\usepackage{graphicx}
\usepackage{natbib}
\usepackage{amsmath}
\usepackage[version=3]{mhchem} 

\shorttitle{Water loss from terrestrial planets}
\shortauthors{Wordsworth et al.}

\begin{document}


\title{Water loss from terrestrial planets with CO$_2$-rich atmospheres}


\author{R. D. Wordsworth}
\affil{Department of the Geophysical Sciences, University of Chicago, 60637 IL, USA}
\email{rwordsworth@uchicago.edu}

\and

\author{R. T. Pierrehumbert}
\affil{Department of the Geophysical Sciences, University of Chicago, 60637 IL, USA}



\begin{abstract}
Water photolysis and hydrogen loss from the upper atmospheres of terrestrial planets is of fundamental importance to climate evolution but remains poorly understood in general. Here we present a range of calculations we performed to study the dependence of water loss rates from terrestrial planets on a range of atmospheric and external parameters. We show that CO$_2$ can only cause significant water loss by increasing surface temperatures over a narrow range of conditions, with cooling of the middle and upper atmosphere acting as a bottleneck on escape in other circumstances. Around G-stars, efficient loss only occurs on planets with intermediate CO$_2$ atmospheric partial pressures (0.1 to 1~bar) that receive a net flux close to the critical runaway greenhouse limit. Because G-star total luminosity increases with time but XUV/UV luminosity decreases, this places strong limits on water loss for planets like Earth. In contrast, for a CO$_2$-rich early Venus, diffusion limits on water loss are only important if clouds caused strong cooling, implying that scenarios where the planet never had surface liquid water are indeed plausible. Around M-stars, water loss is primarily a function of orbital distance, with planets that absorb less flux than $\sim270$~W~m$^{-2}$ (global mean) unlikely to lose more than one Earth ocean of H$_2$O over their lifetimes unless they lose all their atmospheric N$_2$/CO$_2$ early on. Because of the variability of H$_2$O delivery during accretion, our results suggest that many `Earth-like' exoplanets in the habitable zone may have ocean-covered surfaces, stable CO$_2$/H$_2$O-rich atmospheres, and high mean surface temperatures.
\end{abstract}

\section{Introduction}\label{sec:intro}

Understanding the factors that control the water inventories of rocky planets is a key challenge in planetary physics.
In the inner Solar System, surface water inventories currently vary widely: Mars has an estimated 7-20 metres global average H$_2$O
as ice in its polar caps \citep{Morschhauser2011}, Earth has $\sim2.5$~km average H$_2$O as liquid oceans and polar ice caps,
and Venus has only a small quantity ($<0.2$~m global average) in its atmosphere and an entirely dry surface \citep{Chassefiere2012}.
Clearly, these gross differences are due to some combination of variations in the initial inventories and subsequent evolution.

Water is important on Earth most obviously because it is essential to all life, but major uncertainties remain regarding how it was delivered, how it is partitioned between the surface and mantle, and how much has escaped to space over time \citep{Kasting1983,Hirschmann2006,Pope2012}.
Estimating the initial inventory is difficult because water delivery to planetesimals in the inner Solar System during accretion was a stochastic process \citep{Raymond2006,OBrien2006}. However, it appears most likely that Earth's initial water endowment was greater than that of Venus by a factor $\sim3$ or more.

On Venus, surface liquid water may have been present early on but later lost during a H$_2$O runaway or moist stratosphere\footnote{We prefer the term `moist stratosphere'
 to the more commonly used `moist greenhouse' because Earth today is a planet where the
 greenhouse effect is dominated by water vapour.}
phase. In this scenario, large amounts of water would have been dissociated in the high atmosphere by extreme and far ultraviolet (XUV, FUV) photolysis, leading to irreversible hydrogen escape and oxidation of the crust and atmosphere \citep{Kombayashi1967,Ingersoll1969,Kasting1988,Chassefiere1996}. The high ratio of deuterium to hydrogen in the present-day Venusian atmosphere [$\sim$120 times that on Earth; \cite{deBergh1991}] strongly suggests there was once more water present on the planet, but estimating the size and longevity of the early H$_2$O inventory directly from isotope data is difficult \citep{Selsis2007}. It has also been argued based on Ne and Ar isotope data that Venus was never water-rich, and has had high atmospheric CO$_2$ levels since shortly after its formation due to rapid early H$_2$O loss followed by mantle crystallization \citep{Gillmann2009,Chassefiere2012}.

\begin{figure}[H]
	\begin{center}
		{\includegraphics[width=3.5in]{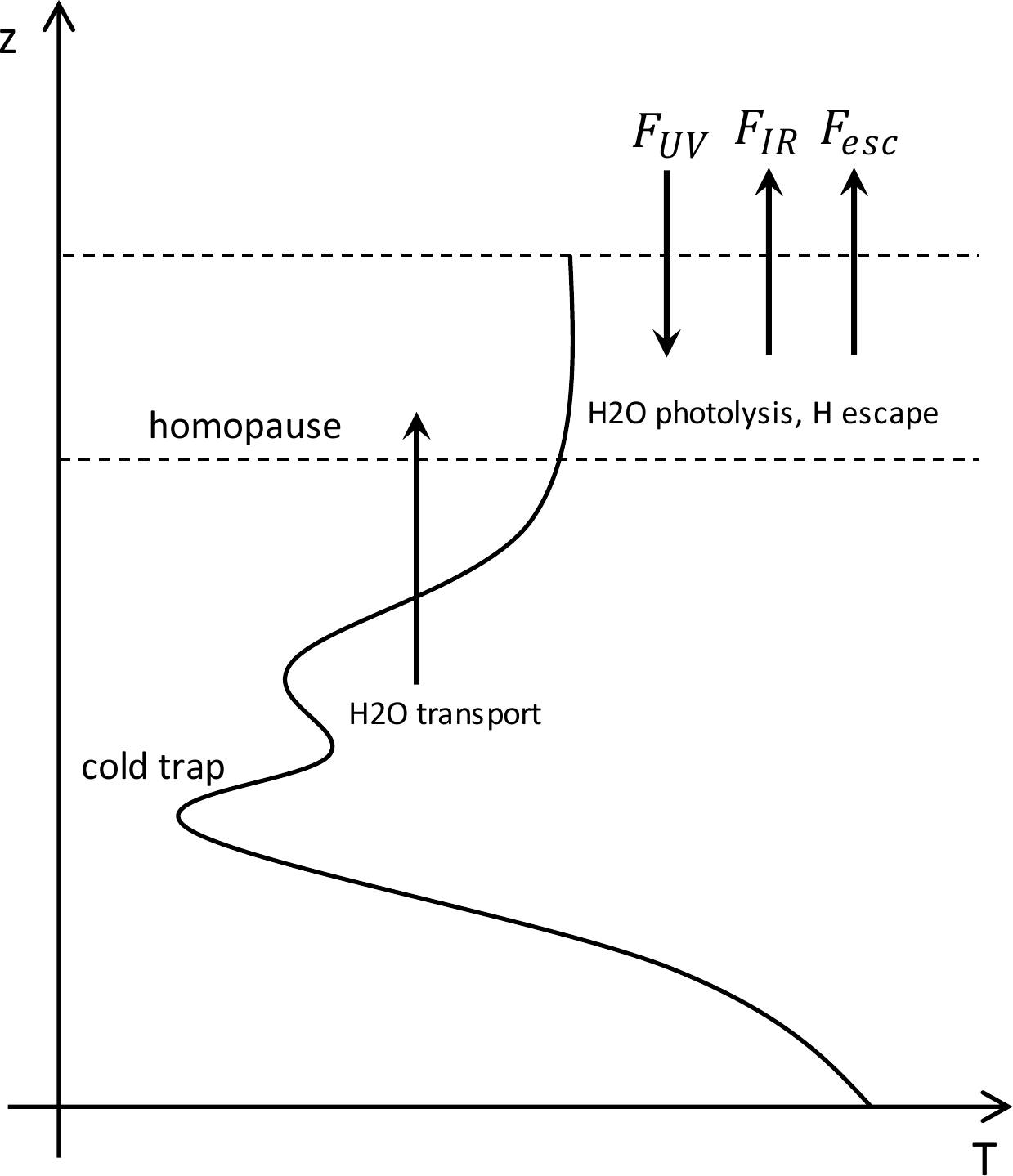}}
	\end{center}
	\caption{Schematic atmospheric temperature profile with the main processes influencing water photolysis and hydrogen loss in terrestrial planetary atmospheres indicated alongside. Transport of H$_2$O from the surface to upper atmosphere is limited by the cold trap. Hydrogen loss rates are controlled by the temperature of the upper atmosphere, which is primarily dependent on a balance between XUV and FUV absorption, IR emission, and the energy carried away by escaping particles.}
	\label{fig:general_schematic}
\end{figure}

For planets with climates that are not yet in a runaway state, the rate of water loss is constrained by the supply of H$_2$O to the high atmosphere. A key factor in this is the temperature of the coldest region of the atmosphere or cold trap, which limits the local H$_2$O mixing ratio by condensation. When cold trap temperatures are low, the bottleneck in water loss becomes diffusion of H$_2$O through the homopause, rather than the rate of H$_2$O photolysis or hydrogen escape to space (Fig.~\ref{fig:general_schematic}).

The extent to which the cold trap limits water loss strongly depends on the amount of
CO$_2$ in the atmosphere. First, CO$_2$ affects the total water content of the atmosphere because it increases surface temperatures by the greenhouse effect. However, the strength of its 15 and 4.3~$\mu$m bands allows efficient cooling to space even at low pressures, so it also plays a key role in determining the cold trap temperature \citep{Pierrehumbert2011BOOK}. Finally, CO$_2$ can also directly limit the escape of hydrogen in the highest part of the atmosphere, because its effectiveness as an emitter of thermal radiation in the IR means it can `scavenge' energy that would otherwise be used to power hydrogen escape \citep{Kulikov2006,Tian2009b}. The history of water on terrestrial planets should therefore be intimately related to that of carbon dioxide.

On Earth, it is generally believed that atmospheric CO$_2$ levels are governed by the crustal carbonate-silicate cycle on geological timescales: increased surface temperatures cause increased rock weathering rates, which increases the rate of carbonate formation, in turn decreasing atmospheric CO$_2$ and hence surface temperature \citep{Walker1981}. Nonetheless, observational studies of silicate weathering rates present a mixed picture. While silicate cation fluxes
in some regions of the Earth (particularly alpine and submontane catchments) are temperature-limited, in other regions (e.g., continental cratons) the rate of physical erosion appears to be the limiting factor \citep{West2005}. The picture is also complicated by basalt carbonization on the seafloor (seafloor weathering). This process is a net sink of atmospheric CO$_2$, but its rate is uncertain and probably only weakly dependent on surface temperature \citep{Caldeira1995,Sleep2001,LeHir2008}.

An accurate understanding of the role of CO$_2$ in the evolution of planetary water inventories will also be important for interpreting future observations of terrestrial\footnote{Throughout this article, we use the term `terrestrial'
to refer to planets of approximate Earth mass (0.1-10~$m_E$) that receive a stellar flux somewhere between that of Venus and Mars and have atmospheres dominated by elements heavier than H and He.} exoplanets.
Because of the diversity of planetary formation histories, it is likely that many terrestrial exoplanets will form with much more H$_2$O than Earth possesses. Depending on the efficiency of processes that partition water between the surface and mantle, many such planets would then be expected to have deep oceans, with little or no rock exposed to the atmosphere \citep{Kite2009,Elkins2011}. Given an Earth/Venus-like total CO$_2$ inventory, these waterworlds\footnote{Here we use the term `waterworld' for a body with enough surface liquid water to prevent subaerial land, but not so much H$_2$O as to inhibit
volatile outgassing [see e.g., \cite{Kite2009,Elkins2011}], following \cite{Abbot2012}. We use the term `ocean planet' for any
planet covered globally by liquid H$_2$O, without any constraint on the total water volume \citep{Leger2004,Fu2010}.} could be expected to have much higher atmospheric CO$_2$ than Earth today due to inhibition of the land carbonate-silicate cycle. \cite{Abbot2012} suggested that waterworlds might undergo `self-arrest', because if large amounts of CO$_2$ were in their atmospheres, they could enter a moist stratosphere state, leading to irreversible water loss via hydrogen escape until surface land became exposed. However, they neglected the effects of CO$_2$ on the middle and upper atmosphere in their analysis.

Even for planets that do have exposed land at the surface, there is currently little consensus as to the extent to which the carbonate-silicate
cycle will resemble that on Earth. Some studies have argued that plate tectonics becomes inevitable as a planet's mass increase, suggesting that in many cases the cycling of CO$_2$ between the crust and mantle, and hence temperature regulation, will be efficient \citep{Valencia2007}. However, other models have suggested that super-Earths may mainly exist in a stagnant-lid regime \citep{ONeill2007}, or that the initial conditions may dominate subsequent mantle evolution \citep{Lenardic2012}. Fascinatingly, some recent work has suggested that the abundance of water in the mantle may be more important to geodynamics than the planetary mass \citep{Korenaga2010,ORourke2012}. Finally, even in the absence of other variations, tidally locked planets around M-stars should have very different carbon cycles from Earth due to the concentration of all incoming stellar flux on the permanent day side \citep{Kite2011,Edson2012}.

In light of all these uncertainties, it seemed clear to us that the role of atmospheric CO$_2$ in evolution of the water inventory deserved to be studied independently of the surface aspects of the problem. To this end, we have performed iterative radiative-convective calculations of the cold-trap temperature and escape calculations that include the scavenging of UV energy by NLTE CO$_2$ cooling, in order to estimate the role of CO$_2$ in water loss via photolysis for a wide range of planetary parameters. Some previous runaway greenhouse calculations tackled the climate  aspects of this problem for the early Earth \citep{Kasting1986,Kasting1988} but assumed a fixed stratospheric temperature. One very recent study, \cite{Zsom2013}, did perform some calculations where the stratospheric temperature was varied, but only in the limited context of investigating the habitability of dry `Dune' planets with low H$_2$O inventories orbiting close to their host stars, following \cite{Abe2011} and \cite{Leconte2013}. An additional motivation for our work was understanding how shortwave absorption affects the atmospheric temperature structure close to the runaway limit. Previous radiative-convective work on this issue simply assumed a moist adiabatic temperature structure in the low atmosphere.

First, we calculate stratospheric saturation using a standard approach with fixed stratospheric temperature and explain the fundamental behaviour of the system via a scale analysis. We then use an iterative procedure to calculate equilibrium temperature and water vapour profiles self-consistently. We show that in certain circumstances, strong temperature inversions may occur in the low atmosphere due to absorption of incoming stellar radiation, which may have important implications for the nature of the runaway greenhouse in general. Taking conservative upper limits on stratospheric H$_2$O levels, we combine the resulting cold-trap H$_2$O diffusion limits with energy-balance escape calculations to estimate the maximum water loss rates as a function of time and atmospheric CO$_2$ content for planets around G- and M-class stars. We then estimate the sensitivity of our conclusions to cloud radiative forcing effects, atmospheric N$_2$ content, surface gravity, and the early impactor flux. In Section \ref{sec:method} we describe our method, in Section \ref{sec:results} we present our results, and in Section \ref{sec:discuss} we discuss the implications for Earth, early Venus, and the evolution and habitability of terrestrial exoplanets.

\section{Method}\label{sec:method}

We perform radiative-convective and escape calculations in 1D, with the implicit (and standard) assumption that heat and humidity redistribution across the planet's surface is efficient and hence a 1D column can be used to represent the entire planet. The uncertainties introduced by this approach are discussed in Section \ref{sec:discuss}. Generally, we assume an N$_2$-H$_2$O-CO$_2$ atmosphere with present-day Earth gravity and atmospheric nitrogen inventory, although we also performed simulations where these assumptions were relaxed. See Table~\ref{tab:params} for a summary of the basic parameters used in the model.

\subsection{Thermodynamics}

The expression used for the moist adiabat is central to any radiative-convective calculation close to the runaway greenhouse limit.
To calculate the saturation vapour pressure and vaporization latent heat of water as a function of pressure, we used the NBS/NRC steam tables \citep{Haar1984,Marcq2012}. We used data from \cite{CRC2000} to calculate analytical expressions for the variation of constant-pressure specific heat capacity $c_{p,i}$ by species $i$ as a function of temperature
\begin{eqnarray}
c_{p,\mbox{N}_2} &=& 1018.7 + 0.078T \quad \mbox{J kg}^{-1}~\mbox{K}^{-1} \\
c_{p,\mbox{CO}_2} &=& 574.8 + 0.875T \quad \mbox{J kg}^{-1}~\mbox{K}^{-1} \\
c_{p,\mbox{H}_2\mbox{O}} &=& 1867.1 - 0.258T +8.502\times10^{-4}T^2 \quad \mbox{J kg}^{-1}~\mbox{K}^{-1},
\label{eq:cp_vary}
\end{eqnarray}
based on a least-squares fit of data between 175 and 600~K. The non-condensible specific heat capacity $c_{p,n}$ was then calculated as a linear combination of $c_{p,\mbox{N}_2}$ and $c_{p,\mbox{CO}_2}$ weighted by volume mixing ratio. The \emph{total} $c_p$, which was calculated with $c_{p,\mbox{H$_2$O}}$ included,
was used to calculate radiative heating rates, and the dry adiabat in convective atmospheric regions where H$_2$O was not condensing.

We related pressure and temperature as
\begin{equation}
\frac{\mbox{d ln }p}{\mbox{d ln }T} = \frac{p_v}{p}\frac{\mbox{d ln }p_v}{\mbox{d ln }T} + \frac{p_n}{p}\left(1 + \frac{\mbox{d ln }\rho_v}{\mbox{d ln }T} - \frac{\mbox{d ln }\alpha_v}{\mbox{d ln }T} \right)\label{eq:adia1}
\end{equation}
with $p_n$ and $p_v$ the partial pressures of the non-condensible and condensible components, respectively,
following \cite{Kasting1988}.
The density ratio $\alpha_v \equiv \rho_v \slash \rho_n$ was related to temperature in the standard way
\begin{equation}
\frac{\mbox{d ln }\alpha_v}{\mbox{d ln }T} = \frac{{R_n \frac{\mbox{d ln }\rho_v}{\mbox{d ln }T} - c_{V,n} - \alpha_v \frac{\mbox{d}s_v}{\mbox{d ln }T} }}{{\frac{\alpha_v L} T + R_n}},\label{eq:adia2}
\end{equation}
with $L$ the latent heat, $s_v$ the entropy of vaporization and $c_{V,n}$, $R_n$ the constant-volume specific heat capacity and specific gas constant, respectively, for the non-condensing component. Although (\ref{eq:adia1}) and (\ref{eq:adia2}) are usually claimed to apply to cases where the condensible component behaves as a non-ideal gas, the starting point for the derivation of (\ref{eq:adia1}) is Dalton's Law, $p=p_n+p_v$ [Eqn. (A1) in \cite{Kasting1988}], which itself requires the implicit assumption that both gases in the mixture are ideal\footnote{It is always true for the
total \emph{number density} that $n=n_n+n_v$, but to relate this to pressure, the ideal gas law is required.}.
A self-consistent derivation of the moist adiabat for a non-ideal condensate would require a non-ideal gas equation for high density N$_2$/CO$_2$ and H$_2$O mixtures. Rather than attempting this in our analysis, we simply treated all gases as ideal, with the exception that we allowed the values of $c_p$ (N$_2$, CO$_2$ and H$_2$O) and $L$ (H$_2$O only) to vary with temperature and pressure. In Section \ref{sec:results}, we demonstrate that
this approximation is unlikely to result in significant errors in our results.

In most simulations, the total mass of N$_2$ in the atmosphere was fixed, the volume mixing ratio of CO$_2$ vs. N$_2$ was varied, and the H$_2$O mixing ratio as a function of pressure calculated from (\ref{eq:adia2}). Because the relationship between the mass column and surface pressure of a given species depends on the local mean molar mass of the atmosphere $\overline \mu$, for a given surface temperature it was necessary to find the correct surface partial pressure of N$_2$ via an iteration procedure at the start of each calculation.

\subsection{Radiative transfer}

For the radiative transfer, a two-stream scheme \citep{Toon1989} combined with the correlated-$k$ method for calculation of gaseous absorption coefficients was used as in previous studies \citep{Wordsworth2010b,Wordsworth2012a}. The HITRAN~2008 database was used to compute high-resolution CO$_2$ and H$_2$O absorption spectra from 10 to 50,000 cm$^{-1}$ using the open-source software \emph{kspectrum}\footnote{https://code.google.com/p/kspectrum/.}.
Kspectrum computes spectral line shapes using the Voigt profile, which incorporates both Lorentzian pressure broadening and Doppler broadening. The latter effect is important at low pressures and high wavenumbers, and must be taken into account for accurate computation of shortwave heating in the high atmosphere.
We produced data on a 14 $\times$ 8 $\times$ 12 temperature, pressure and H$_2$O volume mixing ratio grid of values $T = \{100, 150,\ldots, 750 \}$ K, $p  = \{ 10^{-2}, 10^{-1}, \ldots, 10^5 \} $ mbar and $q_{H_2O}=\{0, 10^{-7}, 10^{-6}, \ldots, 10^{-1}, 0.9, 0.99, 0.999, 1.0 \}$, respectively.

One difficulty in radiative calculations involving high CO$_2$ and H$_2$O is that foreign broadening coefficients in most databases are given with
(Earth) air as the background gas. CO$_2$-H$_2$O line-broadening coefficients do not exist for most spectral lines, and experimental studies have shown that simple scaling of air broadening coefficients is generally too inaccurate to be useful \citep{Brown2007}. To get around this problem, we used the self-broadening coefficients of CO$_2$ and H$_2$O to account for interactions between the gases. This seemed more reasonable than assuming air broadening throughout, because the self-broadening coefficients of both gases are generally greater. The error this introduces in our results is likely to be small compared to larger uncertainties due to e.g., cloud radiative effects (see Section~\ref{subsec:clouds}).

The water vapour continuum was included using the formula in \citet[][pp. 260-261]{Pierrehumbert2011BOOK}, which itself is based on the MT\_CKD scheme \citep{Clough1989}. This scheme includes terms for the self and foreign continua of H$_2$O. The latter is calculated for H$_2$O in terrestrial air and hence may be slightly different at high CO$_2$ levels. However, this is unlikely to affect our results, because the H$_2$O self-continuum dominates the foreign continuum at all wavelengths \citep{Pierrehumbert2011BOOK}. For CO$_2$ CIA, the `GBB' parameterization described in \cite{Wordsworth2010} was used \citep{Gruszka1997,Baranov2004}. Even for moderate surface temperatures, the absorption in the regions where CO$_2$ CIA absorption is strong (0-300~cm$^{-1}$ and 1200-1500~cm$^{-1}$) was dominated by water vapour, so its accuracy was not of critical importance to our results.

Rayleigh scattering coefficients for H$_2$O, CO$_2$ and N$_2$ were calculated using the refractive indices from \citet[][p. 332]{Pierrehumbert2011BOOK}, and the total scattering cross-section in each model layer was calculated accounting for variation
of the atmospheric composition with height.
We considered including the wavelength dependence of the refractive index, as in \cite{vonParis2010}, but existing data appear to have been calculated for present-day Earth conditions only and therefore would have added little additional accuracy. The solar spectrum used was derived from the VPL database \citep{Segura2003}. For the M-star calculations we used the AD~Leo spectrum, as in previous studies \citep{Segura2003,Wordsworth2010b}. In the main calculations, we neglected the radiative effects of clouds and tuned the surface albedo $A_s$ to a value (0.23) that allowed us to reproduce present-day Earth temperatures with present-day CO$_2$ levels. We explore the sensitivity of our results to clouds in Section~\ref{subsec:clouds}. For these calculations, Mie scattering theory was used to compute water cloud optical properties, as in \cite{Wordsworth2010b}. {XUV and UV heating was unimportant to the overall radiative budget of the middle and lower atmosphere even under elevated flux conditions, and hence was only taken into account in the upper-atmosphere escape calculations (next section)}.

Eighty vertical levels were used, with even spacing in log pressure coordinates between $p_{surf}$ and $p_{top}=2$~Pa. In the main simulations, where the stratospheric temperature was not fixed, atmospheric temperatures followed the moist adiabat until radiative heating exceeded cooling, after which temperatures were iterated to local radiative equilibrium (see Section \ref{subsec:inversions} for details). To find global equilibrium solutions [i.e., outgoing longwave radiation (OLR) $-$ absorbed stellar radiation (ASR) = 0], we initially considered using a standard iteration of the type $T_{surf}\to T_{surf} + \epsilon_{conv}\frac{ASR-OLR}{\sigma T_{rad}^3}$,with $T_{rad}=(OLR/\sigma)^{1\slash4}$. However, we found several situations where multiple solutions for $T_{surf}$ and $T(p)$ were possible for the same stellar forcing, due essentially to the fact that CO$_2$ and H$_2$O both have shortwave and longwave effects (see Section~\ref{subsec:inversions}). We therefore performed simulations over a range of $T_{surf}$ values for every simulation, calculated the radiative balance in each case, and then found the equilibrium solution(s) by linear interpolation. While slightly less accurate than an iterative procedure, this approach allowed us much greater control over and insight into the model solutions.

\subsection{Evolution of atmospheric composition}\label{subsec:escapemethod}

To relate our estimates of upper atmosphere H$_2$O mixing ratio to the total water loss across a planet's lifetime, we coupled our radiative-convective calculations to an energy-balance model of atmospheric escape. We chose not simply to refer to existing results from the literature, because we wanted to constrain escape over a wide range of atmospheric and planetary parameters. To get an upper limit on the escape rate of atomic hydrogen, we considered various constraints, starting with the diffusion limit due to the cold trap.

In the diffusion-limited case, the escape rate of hydrogen from the atmosphere is estimated as
\begin{equation}
\Phi_{diff} = b_{H_2O,n} f_{H_2O} \left( H^{-1}_{n} - H^{-1}_{H_2O}\right)\label{eq:difflimescape}
\end{equation}
where
$f_{H_2O}$ is the H$_2$O volume mixing ratio and $H_{H_2O}$ is the scale height of H$_2$O at the homopause. We assume that H$_2$O diffuses and not H$_2$ or H, because most photolysis occurs well above the cold trap\footnote{Calculation of the H$_2$O photodissociation rate J[H$_2$O] from the absorption cross-section data (see Fig.~\ref{fig:CO2_H2O_XUV}) in a representative atmosphere shows rapid decline to low values below a few Pa. This can be compared with typical cold-trap pressures of 100-1000~Pa.}.
$H_n$ is the scale height of the non-condensible mixture (N$_2$ and CO$_2$), and $b_{H_2O,n}$ is the binary diffusion parameter for H$_2$O and N$_2$/CO$_2$ such that
\begin{equation}
b_{H_2O,n} = \frac{b_{H_2O,CO_2}p_{CO_2} + b_{H_2O,air}p_{N_2}}{ p_{CO_2} + p_{N_2} }
\end{equation}
with $p_{N_2}$ ($p_{CO_2}$) the N$_2$ (CO$_2$) partial pressure and $b_{H_2O,CO_2}$ and $b_{H_2O,air}$ calculated using the data given in \cite{Marrero1972}.
The scale heights $H_{H_2O}$ and $H_n$ were calculated using the cold-trap temperature, which was defined as the minimum temperature in the atmosphere (see Section~\ref{sec:results}). The diffusion rate in molecules~cm$^{-2}$~s$^{-1}$ was converted to Earth oceans per Gy assuming total loss of hydrogen and a present-day ocean H$_2$O content of $7.6\times10^{22}$ moles.

\begin{table}[h]
\centering
\caption{Parameters used in the simulations. Standard values are shown in bold.}
\begin{tabular}{ll}
\hline
\hline
Parameter & Values \\
\hline
Stellar zenith angle [degrees] $\theta_z$ & 60.0 \\
Moist adiabat relative humidity  $RH$ & 1.0 \\
Atmospheric nitrogen inventory $M_{N_2}$ [kg~m$^{-2}$] & {$\mathbf {7.8\times10^3}$}, $3.9\times10^4$ \\
Surface albedo $A_s$ & 0.23 \\
Surface gravity $g$ [m~s$^{-2}$] & \textbf{9.81}, 25.0 \\
\hline
\hline
\end{tabular}\label{tab:params}
\end{table}

While our focus was on estimating diffusion limits due to the CO$_2$ cold-trap, we also performed hydrogen escape rate calculations for the
situation where $f_{H_2O}$ approached unity in the upper atmosphere. We investigated limitations due to both the total photolysis rate and the net supply of energy to the upper atmosphere. For the latter, we assumed that the energy balance in the upper atmosphere could be written as
\begin{equation}
F_{UV} = F_{IR} + F_{esc}, \label{eq:upperatmbox}
\end{equation}
where $F_{UV}$ is the ultraviolet (XUV and FUV) energy input from the star, $F_{IR}$ is the cooling to space due to infrared emission, and $F_{esc}$ is the energy carried away by escaping hydrogen atoms created by the photolysis of H$_2$O. Because of the efficiency of H$_2$O and H$_2$ photolysis, H dominates H$_2$ as the escaping species unless the deep atmosphere is reducing, which we assume is not the case here. On a planet with a hydrogen envelope or significant H$_2$ outgassing, H$_2$O photolysis rates would be lower than those we calculate here. For simplicity, we also assume that removal of the excess oxygen from H$_2$O photolysis at the surface is efficient. This is a standard, if somewhat poorly constrained assumption \citep{Kasting1983,Chassefiere2012}. Increased O$_2$ could warm the atmosphere by increasing UV absorption, depending on the level of shielding by H$_2$O. However, O$_2$ can oxidize H before it escapes, and higher levels of atomic oxygen tend to enhance NLTE CO$_2$ cooling \citep{Lopez2001}.
Hence it is unclear how this would affect H escape rates without detailed calculations including photochemistry, which we do not attempt here. {We also neglect the possibility of removal of heavier gases such as CO$_2$ and N$_2$ via XUV heating. This should be a reasonable assumption for all but the most extreme XUV conditions [\cite{Tian2009}, for example, finds that CO$_2$-rich super-Earth atmospheres should be stable for stellar XUV flux ratios below $F_{XUV}\slash F\sim 0.01$].  Depending on stellar activity and the strength of the planet's magnetic field, coronal mass ejection from highly active young stars may also erode substantial quantities of heavy gases from planetary atmospheres \citep{Khodachenko2007,Lammer2007,Lichtenegger2010}. The situation is likely to be most severe for lower mass planets around M-stars, which can lose large amounts of CO$_2$ and N$_2$ if their magnetic moments are weak. In the rest of the paper, we concentrate on hydrogen escape, but we note that in the case of planets in close orbits around M-stars, in particular, our results are contingent on the presence of a sufficiently strong magnetic field to guard against direct loss of the primary atmospheric component.}

For $F_{UV}$, between 10 and 120~nm we used the present-day `medium-activity' spectrum from \cite{Thuillier2004}. This was convolved with wavelength-dependent expressions for evolution of the solar (G-class) XUV flux with time provided in \cite{Ribas2005}, with separate treatment for the Lyman-$\alpha$ peak at 121~nm. Between 120 and 160~nm, a best guess for the UV evolution was used based on \cite{Ribas2010} that yielded an increase to 3$\times$ the present-day level 3.8~Ga. Above 160~nm, we conservatively assumed no change in the UV flux with time. For M-stars, which have inherently more variable XUV emission, we did not attempt to model time evolution, instead using a representative spectrum from a moderately active
nearby M3 dwarf (GJ 436).
For this we used a synthetic combined XUV/UV spectrum provided to us by Kevin France \citep{France2013}. {The XUV portion of this spectrum was normalized using C-III and Lyman-$\beta$ lines (Kevin France, private comm.)}.
In both cases the incoming flux was divided by 4 to account for averaging across the planet, and the contribution of the atmosphere to the planet's
cross-sectional area was neglected.
To calculate absorption by N$_2$, CO$_2$ and H$_2$O and to estimate the H$_2$O photolysis rate, we used N$_2$ and CO$_2$ cross-section data from
\cite{Chan1993b} and \cite{Stark1992}, H$_2$O cross-section data from \cite{Chan1993}, \cite{Fillion2004} and \cite{Mota2005}, and H$_2$O quantum yields from \cite{Huebner1992}.

To calculate the infrared cooling term $F_{IR}$, we used the NLTE `cool-to-space' approximation as in \cite{Kasting1983}. This parameterizes the net volume heating (cooling) rate due to photon emission in the 15~$\mu$m band as
\begin{equation}
q_{CO_2} = n_1 A_{10}\Delta E_{10} \epsilon_{10},\label{eq:NLTE1}
\end{equation}
where $A_{10}$ is the estimated spontaneous emission coefficient for the band,
\begin{equation}
\epsilon_{10} = \frac{1}{1 + \tau\sqrt{2\pi \mbox{ ln}(2.13 + \tau^2)}}
\end{equation}
is the estimated photon escape probability, $\tau = N_{CO_2}\slash 10^{17}$~molec.~cm$^{-2}$, $N_{CO_2}$ is the CO$_2$ column density above a given atmospheric level, $n_1$ is the population of the 1st excited state {and $\Delta E_{10}$ is the energy difference of the ground and excited states}. (\ref{eq:NLTE1}) was integrated numerically over several CO$_2$ scale heights to yield the cooling rate per unit area. Only cooling by the 15~$\mu$m band of CO$_2$ was taken into account. Inclusion of cooling by other CO$_2$ bands or by H$_2$O would have increased our estimate of the IR cooling efficiency and hence decreased our estimates of total water loss in the saturated upper atmosphere limit.

Finally, to find a unique solution to (\ref{eq:upperatmbox}), it was necessary to estimate the escape flux $F_{esc}$ as a function of the temperature at the base of the escaping region, $T_{base}$. For this, we made use of the fact that the escaping form of hydrogen from an atmosphere undergoing water loss should be atomic H, not H$_2$. 
{Atomic hydrogen absorbs hard XUV radiation by ionization at wavelengths below 91~nm with an ionization heating efficiency of 0.15-0.3 \citep{Chassefiere1996,Murray2009}, and has a low collision cross-section, leading to high thermal conductivity \citep{Pierrehumbert2011BOOK}.}
To calculate an upper limit on H escape below the adiabatic blowoff temperature, we assumed a predominantly isothermal flow, with direct XUV-powered escape supplemented by the thermal energy of the H$_2$O and CO$_2$ molecules in the lower atmosphere. For the latter component, we used an analytical expression for the escape flux as a function of $T_{base}$ based on the Lambert $\mathcal W$ function \citep{Cranmer2004}
\begin{equation}
\phi_{hydro}= n_b c_s\sqrt{-\mathcal W_0[-f(r_b\slash r_c)]}
\end{equation}
with $r$ radius,
\begin{equation}
f(x) = x^{-4}\mbox{exp }\left[4\left(1-\frac1x\right)-1\right],
\end{equation}
$r_c =GM_p/(2c_s^2)$ the radius at the transonic point, $G$ the gravitational constant, $M_p$ the planetary mass, $c_s=\sqrt{\gamma_H R_H T_{base}}$ the isothermal sound speed, $\gamma_H$ and $R_H$ the adiabatic index and specific gas constant for atomic hydrogen, and $r_b$ and $n_b$ the radius and density of the base region. We took $M_p=M_E$ (Earth mass) in most cases and assumed $n_b$ to be the total density at the homopause. {In highly irradiated atmospheres, heating can increase the planetary cross-section in the XUV and hence the total amount of radiation absorbed. We crudely account for this effect here by assuming a radius $r_{XUV}=1.3r_E$ for absorption of XUV by H ionization (with $r_E$ Earth's radius). Reference to calculations that account for this effect shows that this is a reasonable assumption for a wide range of XUV forcing values (see e.g., Figs. 3 and 5 of \cite{Erkaev2013}).} The assumption that $n_b$ is the total homopause density overestimates the hydrogen density and hence the total escape rate, but the error due to this is reduced by the fact that the hydrogen scale height is a factor of 18 (44) larger than that of H$_2$O (CO$_2$). As a result, an escaping upper layer of atomic hydrogen may remain in thermal contact with the heavier gases below but decrease in density relatively slowly. We also neglect hydrodynamic drag of these gases on the hydrogen, which again leads us to overestimate escape rates when the incoming UV flux is high.

Finally, to couple the climate and loss rate calculations in time, it was necessary to incorporate the evolution of total stellar luminosity. For this, we assumed no variation for M-stars (constant $L=0.025L_{\sun}$), and evolution for G-stars according to the expression
\begin{equation}
F = F_0\left(1 + \frac 25 \left(1 - t\slash t_{0} \right) \right)^{-1}\label{eq:FYS}
\end{equation}
given in \cite{Gough1981}, with $F_0$ the present day solar flux and $t_0=4.57$~Gy.

\section{Results}\label{sec:results}
\subsection{Variation of OLR and albedo with surface temperature and CO$_2$ mixing ratio}
\begin{figure}[H]
	\begin{center}
		{\includegraphics[width=4.5in]{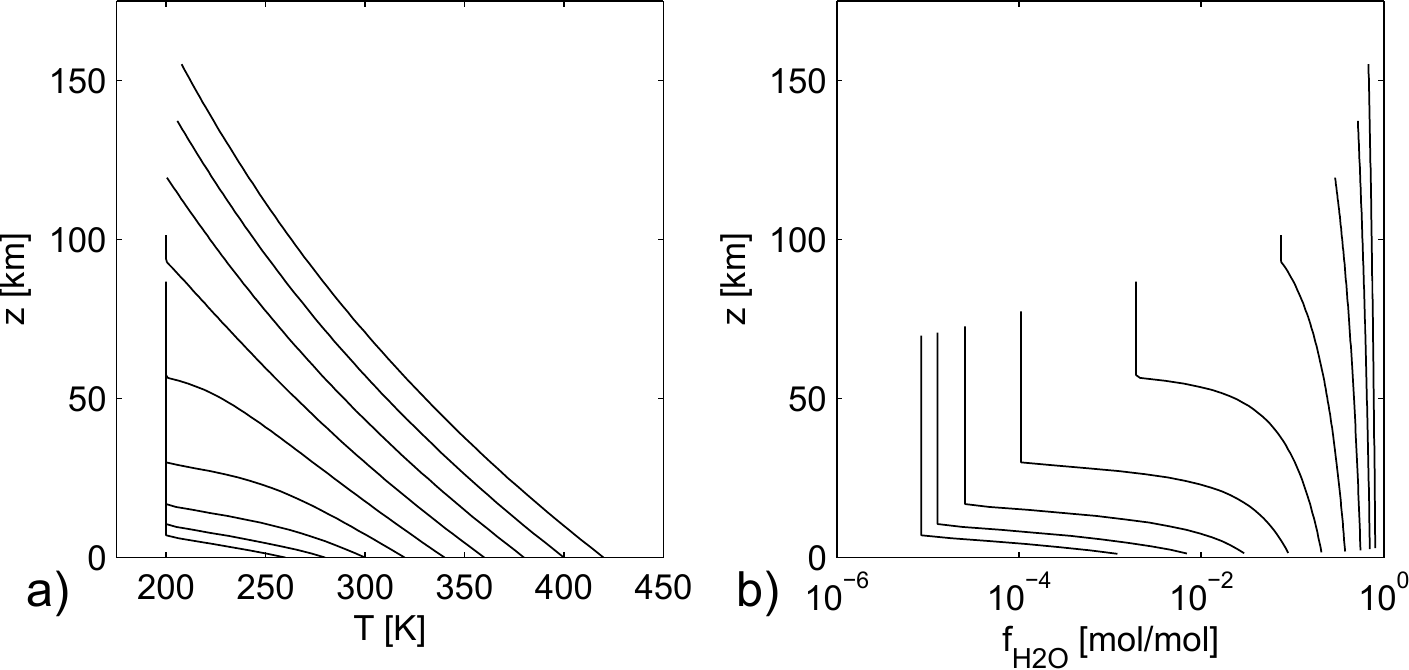}}
	\end{center}
	\caption{a) Temperature and b) H$_2$O volume mixing ratio vs. altitude for tests with fixed stratospheric temperature, 1 bar background N$_2$ and no CO$_2$. Profiles finish at a minimum pressure level $p_{min} = 1$~Pa.}
	\label{fig:kastcompare}
\end{figure}

We first compared the results of our model with the classical runaway greenhouse calculations of \cite{Kasting1988}. For this we assumed 1~bar of N$_2$ as the background incondensible gas and a constant stratospheric temperature of 200~K. Figure \ref{fig:kastcompare} shows the temperature profiles and H$_2$O volume mixing ratios obtained. The results are almost identical to those in \cite{Kasting1988}, demonstrating that the inclusion of the non-ideality terms discussed in Section~\ref{sec:method} makes little difference to the results for this range of surface pressures. Computing the outgoing longwave radiation for this set of profiles, we found a peak of 296~W~m$^{-2}$, compared with $\sim 310$~W~m$^{-2}$ in \cite{Kasting1988} (results not shown). This is close to the value reported in \cite{Pierrehumbert2011BOOK}, which is unsurprising because the H$_2$O continuum dominates the OLR in the runaway limit.\footnote{Note that in \cite{Kopparapu2013}, it is stated that differences between the BPS and CKD continuua  \citep{Shine2012,Clough1989} can cause up to 12~W~m$^{-2}$ difference in the OLR in the runaway limit. However, these authors later claim that their results closely correspond to Fig.~4.37 in \cite{Pierrehumbert2011BOOK}, which was itself calculated using a continuum parameterization based on CKD.
{Alternatively, the differences found vs. line-by-line results in \cite{Kopparapu2013} may be due to line shape assumptions (R. Ramirez, private comm.)}. Nonetheless, a systematic intercomparison between the various continuum schemes for H$_2$O would probably be a useful future exercise.}.

\begin{figure}[H]
	\begin{center}
		{\includegraphics[width=4.5in]{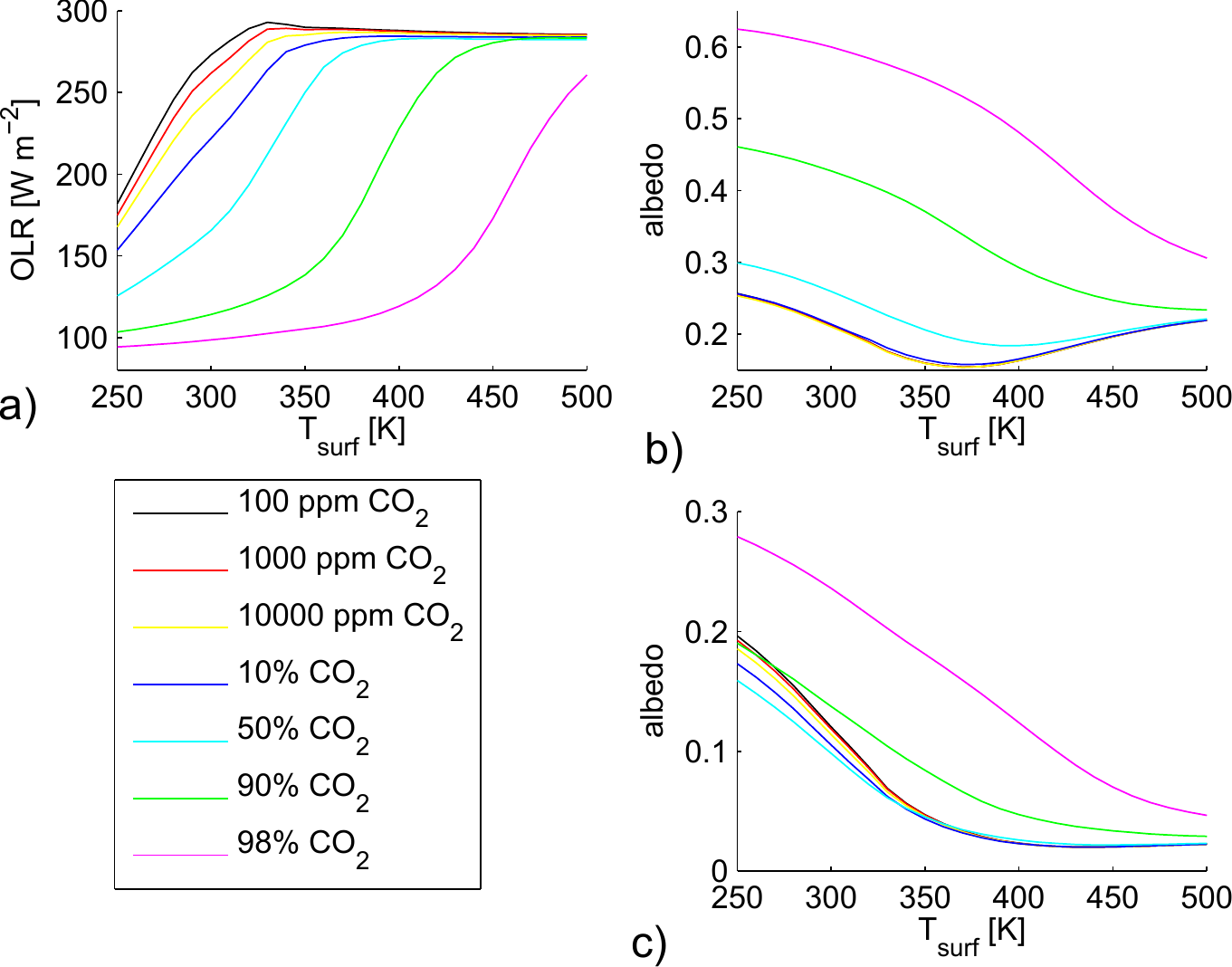}}
	\end{center}
	\caption{a) OLR as a function of surface temperature for various CO$_2$ dry volume mixing ratios, with fixed $T_{strat}=200$~K, 100\% relative humidity, and Earth gravity and present-day atmospheric nitrogen inventory. b, c) Corresponding planetary albedo for G- and M-star incident spectra, respectively.}
	\label{fig:OLR_ALB_vs_Ts_fCO2}
\end{figure}

Next, we calculated the OLR and albedo vs. surface temperature for a range of CO$_2$ dry volume mixing ratios. Figure~\ref{fig:OLR_ALB_vs_Ts_fCO2} shows a) the OLR and b,c) albedo for G- and M-star spectra, respectively, assuming Earth's gravity and present-day atmospheric nitrogen inventory. For intermediate surface temperatures, carbon dioxide reduces the OLR, but by $T_{surf}\sim 500$~K, the runaway limit is approached by all cases except the 98\%~CO$_2$/2\%~N$_2$ atmosphere.
{At high temperatures, the limiting OLR varies between 285.5~W~m$^{-2}$ (100 dry ppm CO$_2$) and 282.5~W~m$^{-2}$ (50\% dry CO$_2$). This is in close agreement with the line-by-line calculations of \cite{Goldblatt2013}; use of the HITEMP 2010 database for H$_2$O would probably have resulted in a reduction in our limiting OLR by a few W~m$^{-2}$. }

CO$_2$ also has a important effect on the planetary albedo, particularly in the G-star case, with a stronger influence at higher temperatures than for the OLR. This can be explained by the fact that all the atmospheres are more opaque in the infrared than in the visible, so CO$_2$ continues to affect the visible albedo even at high temperatures, when the H$_2$O column amount becomes extremely high.

Our planetary albedo values are systematically lower than those in \cite{Kasting1988}, as was also found by \cite{Kopparapu2013} in their recent (cloud-free) revision of the inner edge of the habitable zone. This is caused by atmospheric absorption of H$_2$O in the visible, due to vibrational-rotational bands that were poorly constrained when the radiative-convective calculations in \cite{Kasting1988} were performed, but are included in the HITRAN~2008 database \citep{Rothman2009}. The effects of this absorption beyond simple changes in the planetary albedo are discussed in detail in the next section.

\subsection{Shortwave absorption and low atmosphere temperature inversions}\label{subsec:inversions}

\begin{figure}[H]
	\begin{center}
		{\includegraphics[width=4.5in]{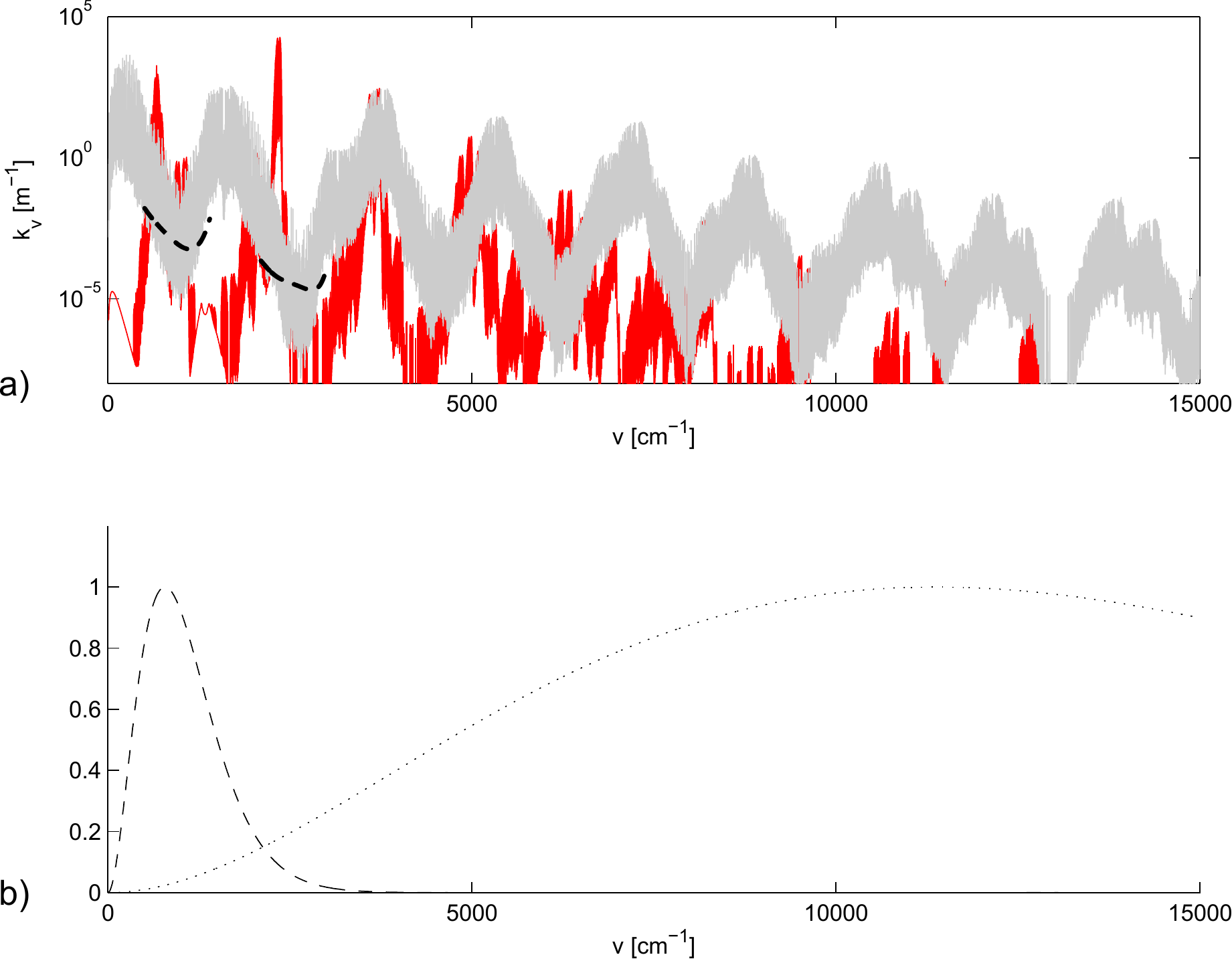}}
	\end{center}
	\caption{a) High-resolution absorption data for CO$_2$ (red) and H$_2$O (gray) used to create correlated-$k$ coefficients for the radiative transfer calculations. Data shown are for pure gas absorption at 400~K and 0.1~bar. The H$_2$O continuum [as defined in \cite{Pierrehumbert2011BOOK}] is indicated by the dashed black lines. b) Normalised blackbody emission at $T = 400$~K and 5800~K (dashed and dotted lines, respectively).}
	\label{fig:spectra}
\end{figure}

\begin{figure}[h]
	\begin{center}
		{\includegraphics[width=6in]{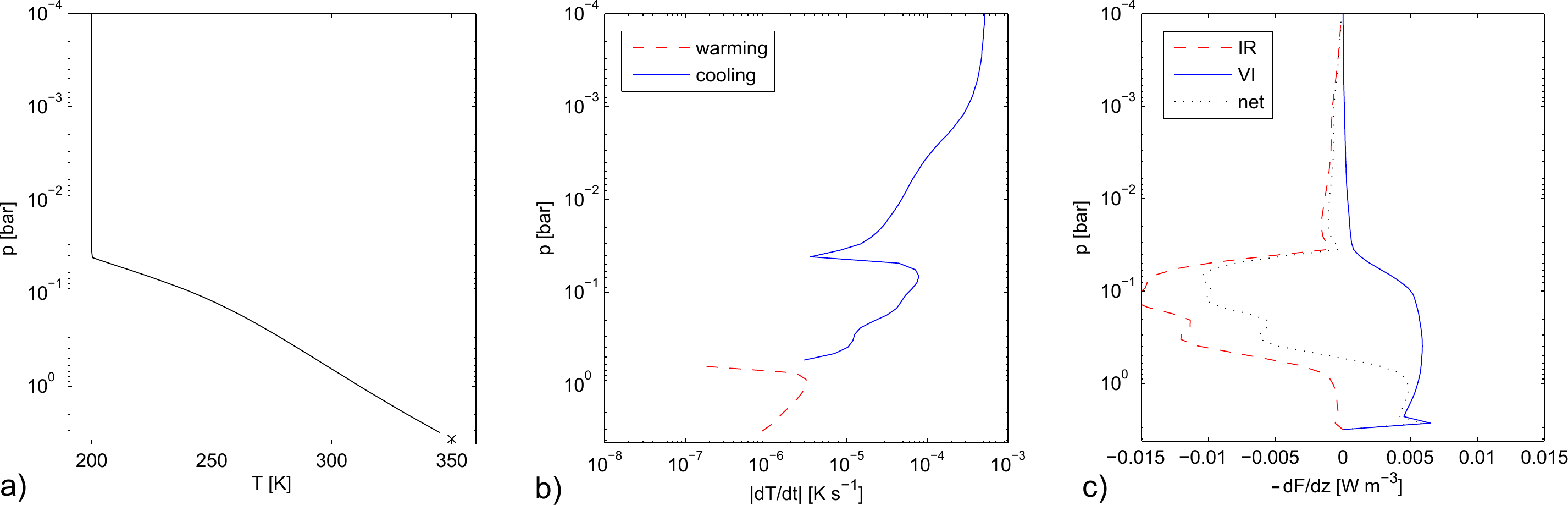}}
	\end{center}
	\caption{a) Temperature profile, b) radiative heating rates and c) flux gradients for an atmosphere with Earth's present day N$_2$ inventory, CO$_2$ dry volume mixing ratio of 0.7, solar forcing of 0.85$F_0$, RH=1.0, $T_{surf}=350$~K and fixed $T_{strat}=200$~K. The curves in b) and c) are related by $\frac{dT}{dt} = -\frac{dF}{dz}\slash \rho c_p$, with $\rho$ and $c_p$ as defined in the main text.}
	\label{fig:radheat}
\end{figure}

The absorption spectra of CO$_2$ and H$_2$O from the far-IR to 0.67~$\mu$~m are shown in Fig.~\ref{fig:spectra} a). For comparison, blackbody curves at 400 and 5800~K are shown in Fig.~\ref{fig:spectra} b). As can be seen, the absorption bands of both gases extend well into the visible spectrum. As a result, when a terrestrial planet's atmospheric CO$_2$ content is high, the amount of starlight reaching the surface is greatly reduced.
When the atmosphere is thick enough, this can qualitatively change the net radiative heating profile in the atmosphere. In Fig.~\ref{fig:radheat}, the temperature profile, radiative heating rates and flux gradients are plotted for a planet with Earth-like gravity and atmospheric N$_2$ inventory, CO$_2$ dry volume mixing ratio of 0.7, $T_{surf}=350$~K and fixed $T_{strat}=200$~K, irradiated by a G-class (Sun-like) star. As can be seen, the visible absorption by CO$_2$ and H$_2$O is strong enough to cause net heating, rather than cooling, in the lower atmosphere.

\begin{figure}[h]
	\begin{center}
		{\includegraphics[width=6in]{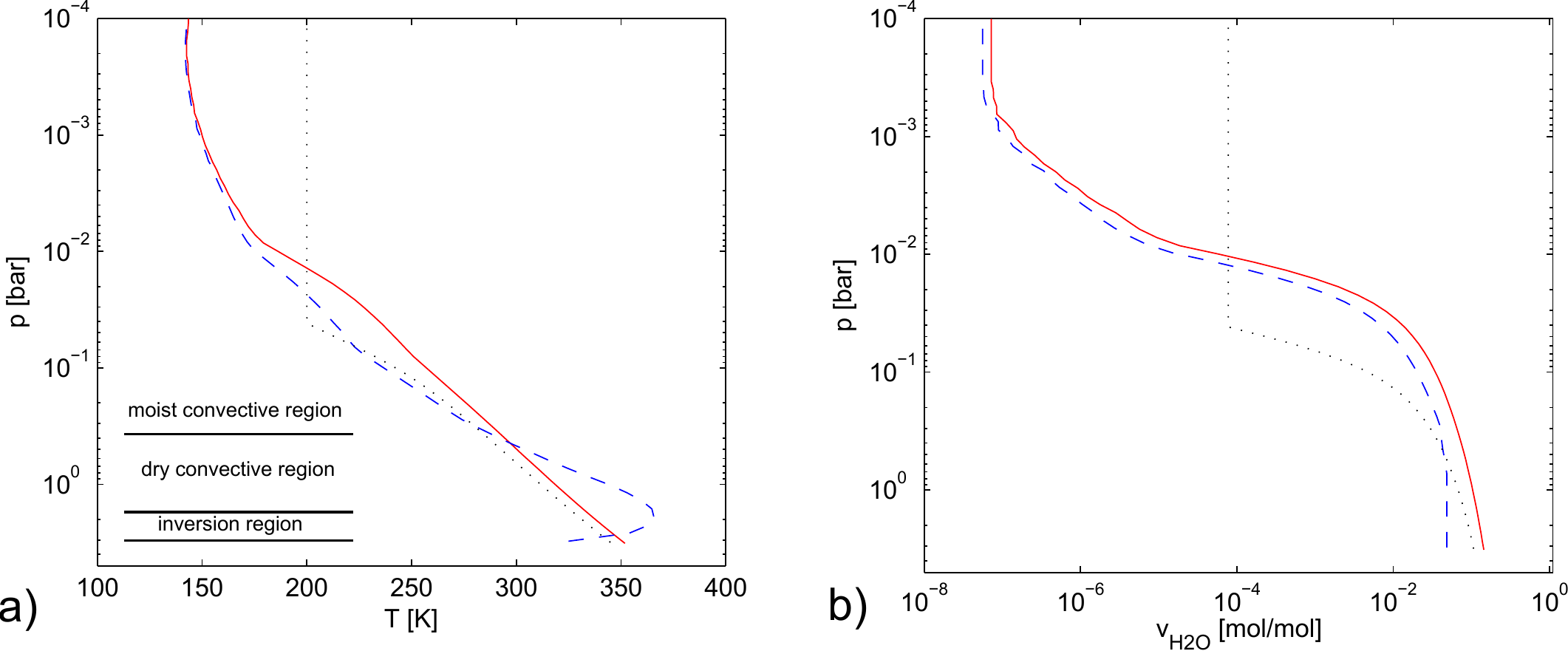}}
	\end{center}
	\caption{a) Temperature profiles and b) H$_2$O volume mixing ratios for the same atmospheric composition as in Fig.~\ref{fig:radheat}. Red solid (blue dashed) curves are for cases where departure from the moist adiabat was inhibited (permitted) in the low atmosphere (below 0.2~bar). The black horizontal lines and text on the left indicate atmospheric regions for the blue dashed curve.}
	\label{fig:radequi}
\end{figure}

To examine the effect of this heating on the atmospheric temperature profiles, we ran the radiative-convective model in time-stepping mode until a steady state was reached (Fig.~\ref{fig:radequi}). In one simulation, we allowed the atmosphere to evolve freely (red line), while in another, we forced the temperature profile to match the moist adiabat below 0.2~bar. For this example, cloud effects were neglected in the calculation of the visible albedo. As can be seen, in both iterative cases, CO$_2$ cooling in the high atmosphere reduces stratospheric temperatures to around 150~K, significantly decreasing $f_{H_2O}$ there. This effect is discussed further in the next section. In addition, in the freely iterative case, the low atmospheric absorption causes a strong temperature inversion to form near the surface. Above the inversion region, the atmosphere again becomes convectively unstable, following the dry adiabat as pressure decreases until the air is once again fully saturated, after which the model returns the temperature profile to the moist adiabat. The resulting reduction in surface temperature and lowered relative humidity ($RH$) in the inversion layer causes $f_{H_2O}$ to decrease slightly more in the upper atmosphere compared to the case where the lower atmosphere was forced to follow a moist adiabatic temperature profile.

We chose this high-CO$_2$ example to give a clear demonstration of the phenomenon, but
this pattern of cooling in the mid-atmosphere but heating at depth should be an inevitable
feature of very moist atmospheres around main sequence stars. Because the H$_2$O continuum
region between 750 and 1200~cm$^{-1}$ (see Fig.~\ref{fig:spectra}) is the ultimate limiting
factor on cooling to space when H$_2$O levels are high, the peak region of IR cooling becomes
fixed around 0.1~bar (see Fig.~\ref{fig:radheat}) once the atmosphere is sufficiently moist.
However, absorption by both H$_2$O and CO$_2$ is weaker per unit mass in the shortwave than in
the longwave, so most stellar absorption
must occur deep in the atmosphere, where the high IR opacity means that radiative cooling
rates are low. Low atmosphere heating is generally even stronger around M-stars than G-stars,
because the red-shift in the stellar radiation increases absorption and decreases the importance of Rayleigh scattering
\citep[e.g., ][]{Kasting1993,Wordsworth2010b}.

Further understanding of the inversion behaviour can be gained by considering the surface energy budget. Because some radiation still reaches the ground even at very high CO$_2$ levels, evaporation of H$_2$O from the surface will still occur whenever a surface liquid source is present. In fact, when the temperature of the low atmosphere is higher than that of the ground, evaporation must increase, as evidenced by the equilibrium surface energy equation
\begin{equation}
F_L = F^{sw}_{abs} + c_p\rho_a C_D \|{\bf v}_{\rm a}\| (T_a - T_{surf}) + \sigma \left(T_a^4 - T_{surf}^4\right).\label{eq:surfbudget}
\end{equation}
Here $F^{sw}_{abs}$ is the incident shortwave radiation from above absorbed by the surface, $C_D$ is a drag coefficient and $\rho_a$, $T_a$ and $\|{\bf v}_{\rm a}\|$ are the atmospheric density, temperature, and mean wind speed near the surface, and it is assumed that the lower atmosphere is optically thick in the infrared. Clearly, $F_L$, the latent heat flux due to evaporation, must be positive to balance the right hand side if $T_a>T_{surf}$. This immediately implies a net loss of mass from the surface as liquid is converted to vapour.

In a steady state, the mass loss due to evaporation must be balanced by precipitation. However, in the regions where the atmosphere undergoes net  radiative heating, $RH$ drops below unity, so evaporation of precipitation should lead to a mass imbalance in the hydrological cycle and hence a net increase in the atmospheric mass over time. If the atmosphere were well-mixed everywhere, this process would continue until the temperature inversion was removed and moist convection could presumably again occur in the low atmosphere.

In reality, the picture is more complex, because convective and boundary layer processes lead to frequent situations where $RH$ varies significantly even on small scales \citep{Pierrehumbert2007}. The large-scale planetary circulation is also important: on the present-day Earth, broad regions of downwelling in the descending branches of the Hadley cells have $RH$ well below 1.0. Interestingly, recent general circulation model (GCM) simulations of moist atmospheres near the runaway limit have also shown evidence of temperature inversions, although so far only for the special case of tidally locked planets around M-stars \citep{Leconte2013}.

In the following analysis, to bracket the uncertainty in the results, we show cases where the atmosphere was allowed to evolve freely alongside those where the atmosphere was forced to follow the moist adiabat in the lower atmosphere. For the latter simulations, we simply switched off shortwave heating deeper than a given pressure (here, 0.2~bar), while still requiring balance between net outgoing and incoming radiation at the top of the atmosphere in equilibrium. As will be seen, the surface temperatures and cold-trap H$_2$O mixing ratios tended to be lower when the atmosphere evolved freely. The general issue of atmospheric temperature inversions due to shortwave absorption in dense moist atmospheres is something that we plan to investigate in more detail in future using a 3D model. It is likely to be particularly important for planets around M-stars, which have elevated atmospheric absorption due to their red-shifted stellar spectra.

\subsection{Dependence of upper atmosphere H$_2$O mixing ratio on CO$_2$ levels}\label{subsec:dependence}

Before performing iterative calculations of the cold-trap temperature, we first calculated the dependence of the upper atmospheric H$_2$O mixing ratio on CO$_2$ levels assuming a fixed (high) stratospheric temperature of $T_{strat}=200$~K. This is close to the skin temperature for Earth today: assuming an albedo of 0.3, $T_{skin}=2^{-1\slash4}\left(OLR\slash\sigma\right)^{1\slash4}=214$~K. However, as should be clear from the iterated profiles in Figure~\ref{fig:radequi}, it represents a considerable overestimate when the main absorbing gas in the atmosphere is non-gray. As previously mentioned, carbon dioxide is particularly effective at cooling the high atmosphere because its strong 15~$\mu$m absorption band remains opaque even at low pressures.

\begin{figure}[h]
	\begin{center}
		{\includegraphics[width=4.5in]{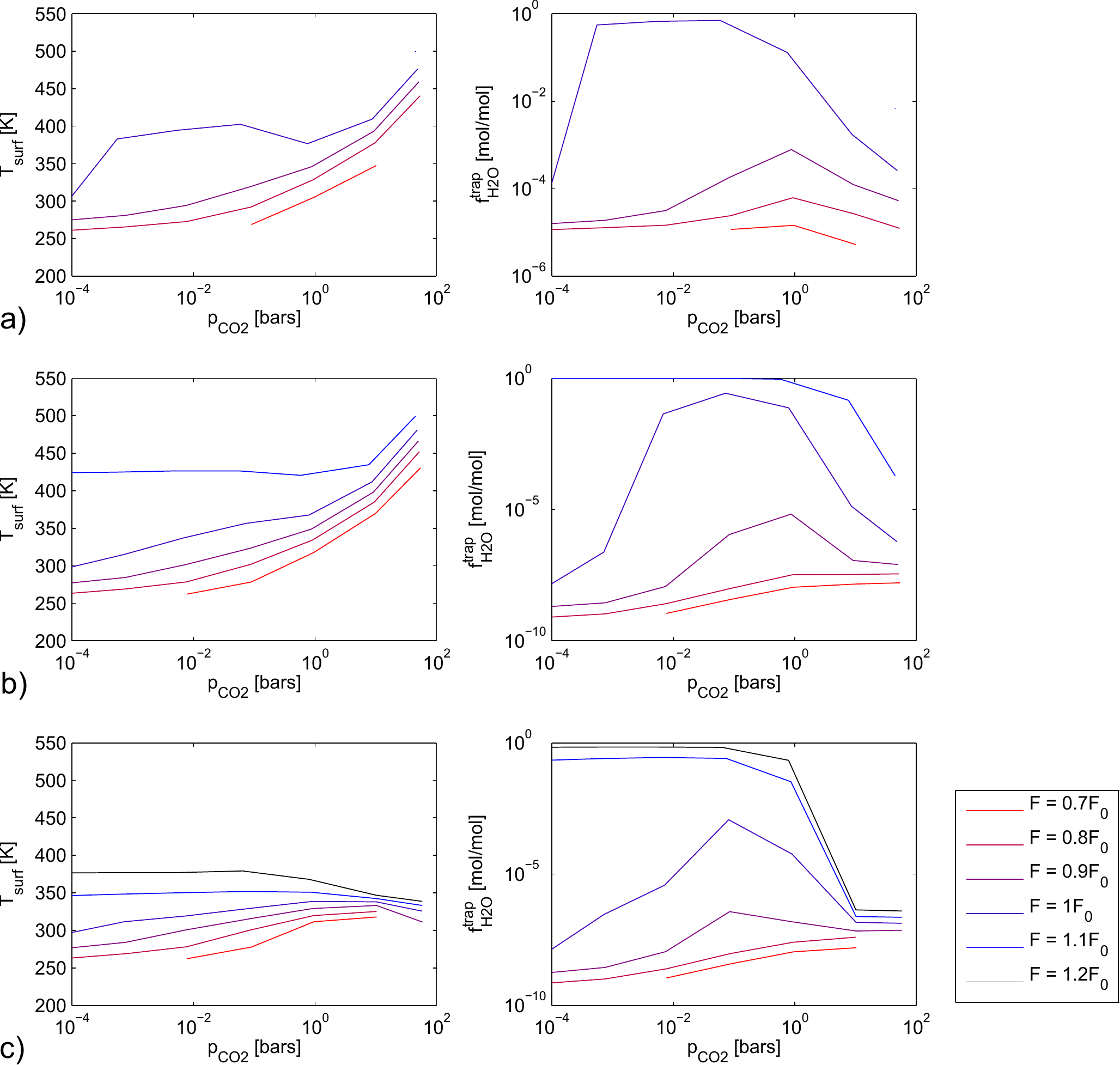}}
	\end{center}
	\caption{ (left) Surface temperature and (right) cold-trap H$_2$O volume mixing ratio as a function of surface CO$_2$ partial pressure for a range of incident solar fluxes. Cases a-c) are for simulations where a fixed stratospheric temperature of 200~K was assumed, where the temperature profile was fixed below 0.2~bar but evolved freely above, and where the entire atmospheric temperature profile evolved freely, respectively. In the latter case, strong temperature inversions formed near the surface due to shortwave H$_2$O absorption.}
	\label{fig:vstrat}
\end{figure}

\begin{figure}[h]
	\begin{center}
		{\includegraphics[width=3.5in]{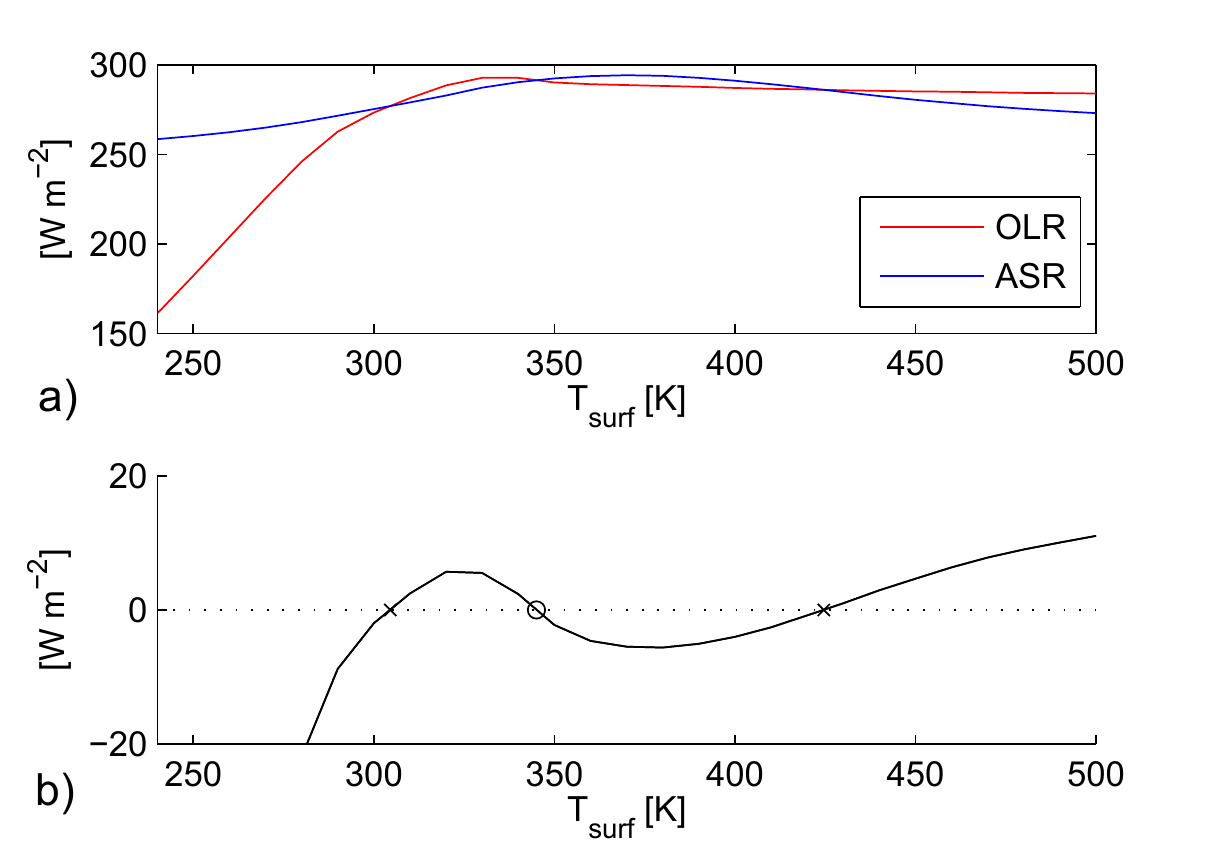}}
	\end{center}
	\caption{a) OLR, ASR and b) OLR - ASR for an atmosphere with three thermal equilibria (two stable solutions shown by crosses, one unstable solution shown by the circle). For this example, $F = 1.025 F_0$ and the CO$_2$ dry volume mixing ratio was 100~ppm.}
	\label{fig:multiple_equilib}
\end{figure}

Figure~\ref{fig:vstrat}a) (left) shows the surface temperature as a function of CO$_2$ surface partial pressure $p_{CO_2}$, for a range of solar forcing values, assuming the planet is Earth and the star is the Sun. When the incoming solar radiation was close to the runaway limit, multiple equilibria were found for some $p_{CO_2}$ values. This was due to the varying behaviour of OLR and albedo with temperature (see Fig.~\ref{fig:multiple_equilib}). In the following analysis, we take the hottest stable solution whenever multiple equilibria are present, in keeping with our aim of a conservative upper limit on stratospheric moistening.

\begin{figure}[h]
	\begin{center}
		{\includegraphics[width=3.5in]{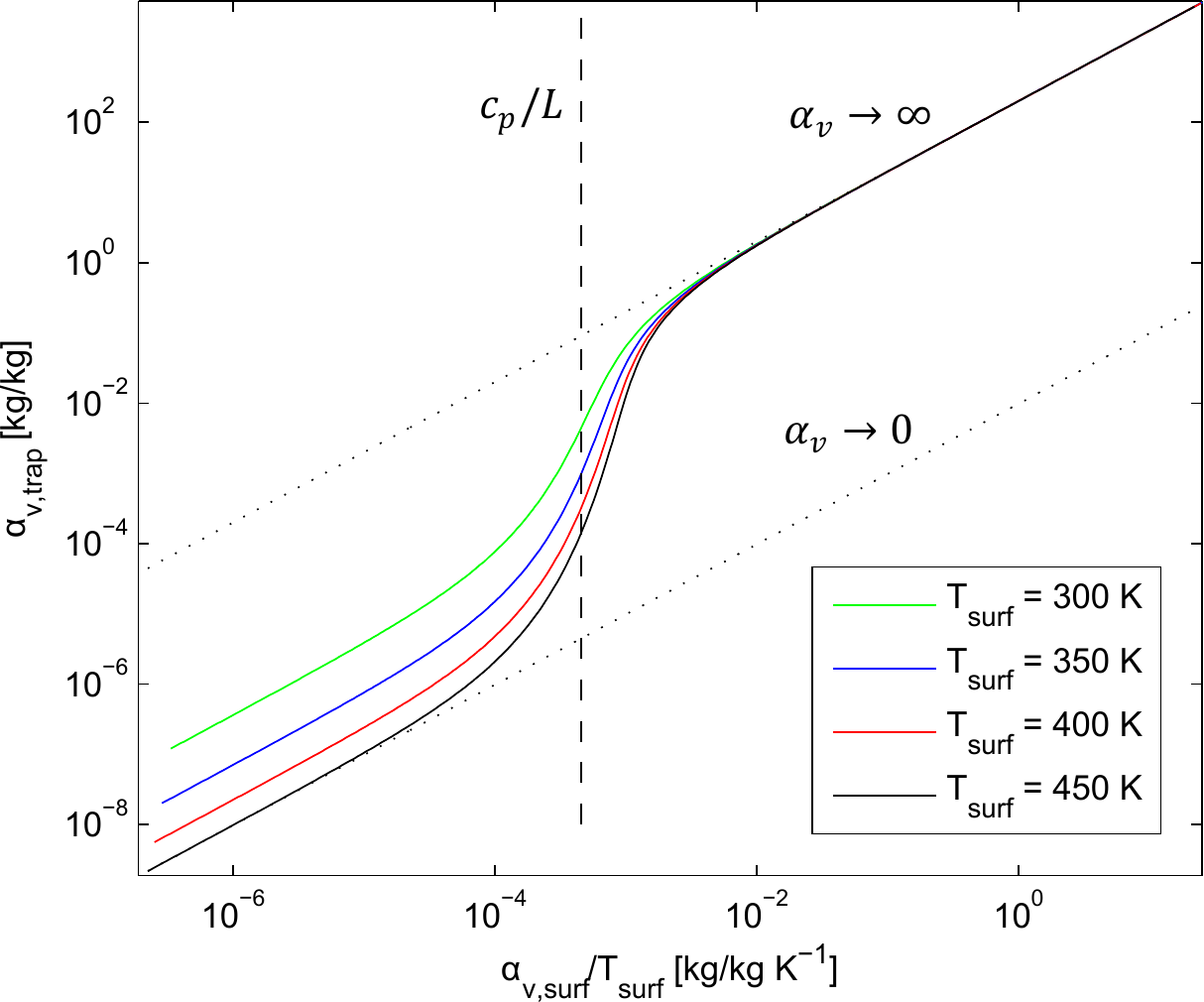}}
	\end{center}
	\caption{Condensible to non-condensible mass mixing ratio at the cold-trap $\alpha_{v,trap}$ vs. $\alpha_{v,surf}/T_{surf}$ according to (\ref{eq:adia_simp1}), for a range of $T_{surf}$ values and $T_{strat}=200$~K. The transition to a moist stratosphere occurs for a narrow range of $\alpha_{v,surf}/T_{surf}$ values centered around $c_{p,n}\slash L$ (dashed line). The behaviour of $\alpha_{v,trap}$ in the limits  $\alpha_{v,surf}\to\infty$ and  $\alpha_{v,surf}\to0$ (for $T_{surf}=450$~K) are shown by the dotted lines.}
	\label{fig:analytic_alpha}
\end{figure}

Figure~\ref{fig:vstrat}a) (right) shows the corresponding mixing ratio of H$_2$O at the cold-trap $f_{H_2O}^{trap}$ for the same range of cases, with the high $T_{surf}$ solution chosen when multiple equilibria were present. As can be seen, increasing CO$_2$ initially moistens the atmosphere at the cold trap by increasing surface temperature. This effect continues until $p_{CO_2}\sim0.1$~bar, after which the cold-trap fraction of H$_2$O declines again, despite the continued increase in surface temperature. Deeper insight into this phenomenon can be gained by studying a semi-analytical model. Equation (\ref{eq:adia2}) can be simplified in the ideal gas, constant $L$ and $c_{V,n}$ limit to
\begin{equation}
\frac{\mbox{d ln }\alpha_v}{\mbox{d }T} = \frac{\left(\alpha_v + \epsilon \right)\slash T - c_{p,n}\slash L }{\alpha_v + R_nT\slash L},\label{eq:adia_simp1}
\end{equation}
assuming that the relationship between $\rho_v$ and $T$ is given by the Clausius-Clayperon equation. Here $\epsilon=m_v \slash m_n$ is the molar mass ratio of the condensing and non-condensing atmospheric components.  In the limit $\alpha_v \to \infty$, (\ref{eq:adia_simp1}) is trivially integrated from the surface to cold trap to yield
\begin{equation}
\frac{\alpha_{v,trap}}{\alpha_{v,surf}} \sim \frac{T_{trap}}{T_{surf}}  \label{eq:adia_limitA}
\end{equation}
Conversely, in the limit $\alpha_v \to 0$, (\ref{eq:adia_simp1}) integrates to
\begin{equation}
\frac{\alpha_{v,trap}}{\alpha_{v,surf}} \sim \mbox{exp}\left[{+\frac{L}{R_v}\left(T_{surf}^{-1} - T_{trap}^{-1}\right)}\right] \left(\frac{T_{surf}}{T_{trap}} \right)^{c_{p,n}\slash R_n} . \label{eq:adia_limitB}
\end{equation}
For temperature ranges and $L$, $c_p$ values appropriate to H$_2$O condensation in an N$_2$/CO$_2$ atmosphere, the transition between these two limits occurs rapidly over a small range of $\alpha_v$ values. Figure~\ref{fig:analytic_alpha}~a) shows $\alpha_{v,trap}$ as a function of $\alpha_{v,surf}$ given $T_{surf}=350$~K and $T_{trap}=150$~K in a pure N$_2$ atmosphere. As can be seen, $\alpha_{v,trap}$ only deviates from the lower and upper limits in a relatively narrow
region. With reference to (\ref{eq:adia_simp1}), we can define a dimensionless moist saturation number
\begin{equation}
\mathcal M \equiv \frac{\rho_{v,surf} L}{\rho_{n,surf} c_{p,n} T_{surf}}.\label{eq:Mdef}
\end{equation}
based entirely on surface values.
As Fig.~\ref{fig:analytic_alpha} shows, the transition to a regime where the upper atmosphere is moist occurs when $\mathcal M>1$. Equivalently, saturation of the upper atmosphere becomes inevitable once the latent heat of the condensible component at the surface exceeds the sensible heat of the non-condensing component. Because of the nonlinearity of the transition between the two regimes, this general scaling analysis can still be used as a guide even when $T_{trap}$ varies, although for quantitative estimates of $\alpha_{v,trap}$ near $\mathcal M=1$, numerical calculations are required, as should be clear from Fig.~\ref{fig:vstrat}.

\begin{figure}[h]
	\begin{center}
		{\includegraphics[width=3.5in]{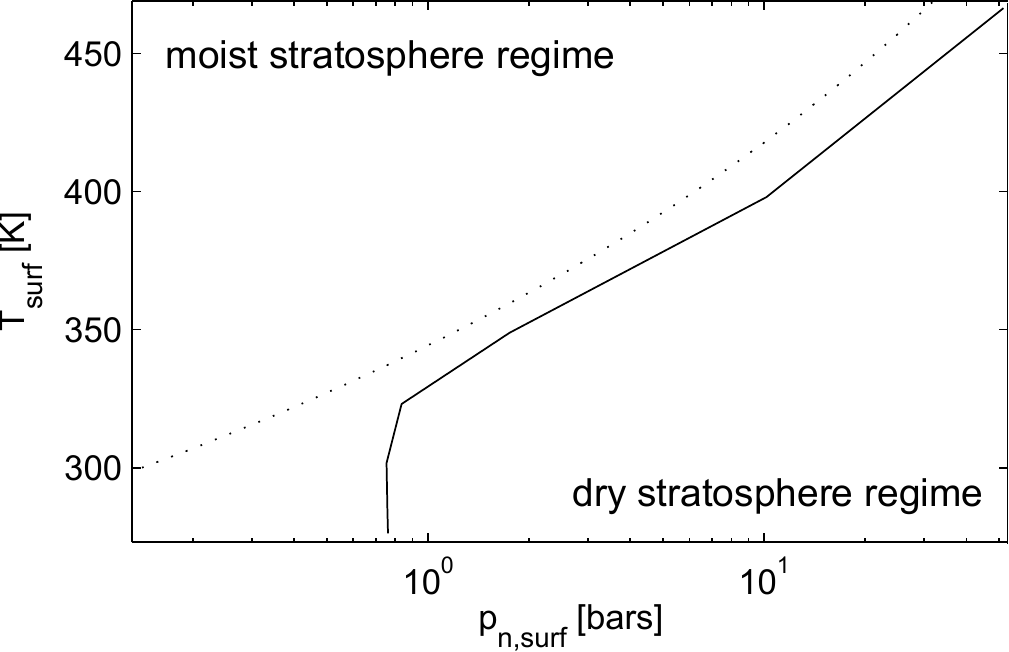}}
	\end{center}
	\caption{Surface temperature as a function of non-condensible surface partial pressure $p_{n,surf}$ (N$_2$ and CO$_2$)
given a solar flux $F=0.9F_0$ (solid line) and assuming fixed $T_{strat}=200$~K. The dashed line shows the $\mathcal M=1$ temperature limit derived from (\ref{eq:adia_Tlimit}). The initial rapid increase of $T_{surf}$ with $p_{n,surf}$ occurs due to the addition of CO$_2$ in small quantities to an initially N$_2$-dominated atmosphere.}
	\label{fig:analytic_Tlimit}
\end{figure}

Assuming a saturated moist adiabat, this definition of $\mathcal M$ allows us to derive an expression for the rate at which $T_{surf}$ must increase with $p_{CO_2,surf}$ in order for the upper atmosphere to remain moist. Given
\begin{eqnarray}
p_{n,surf} &=& \epsilon \frac{p_{sat}(T_{surf})}{\alpha_{v,surf}} \\
           &=& \frac{\epsilon L p_{sat}(T_{surf})}{c_{p,n} T_{surf} \mathcal M}
\end{eqnarray}
then
\begin{equation}
p_{n,surf}(\mathcal M=1) = \frac{\epsilon L p_0}{c_{p,n}}\frac{\mbox{exp}\left[{-\frac{L}{R_v}\left(T_{surf}^{-1} - T_{0}^{-1}\right)}\right]}{T_{surf}} \label{eq:adia_Tlimit}
\end{equation}
and hence
\begin{equation}
p_{CO_2,surf}(\mathcal M=1) =  \frac{\epsilon L p_0}{c_{p,n}}\frac{\mbox{exp}\left[{-\frac{L}{R_v}\left(T_{surf}^{-1} - T_{0}^{-1}\right)}\right]}{T_{surf}}  - p_{N_2,surf}.
\end{equation}
The curve described by (\ref{eq:adia_Tlimit}) is plotted in Fig.~\ref{fig:analytic_Tlimit} alongside the actual increase of temperature with $p_{n,surf}$ for a simulation with $F=0.9F_0$ and variable CO$_2$, for comparison. When CO$_2$ is a minor component of the atmosphere, its greenhouse effect per unit mass is high, so increasing its mixing ratio raises surface temperatures but barely affects $p_{n,surf}$. However, once CO$_2$ is a major constituent, it begins to significantly contribute to $p_{n,surf}$ and hence to the sensible heat content of the atmosphere. In addition, it begins to increases the planetary albedo via Rayleigh scattering (Fig.~\ref{fig:OLR_ALB_vs_Ts_fCO2}b). Then, the increase of $T_{surf}$ with $p_{n,surf}$ is no longer sufficient to allow the climate to cross over into the moist stratosphere regime, and the H$_2$O mixing ratio in the upper atmosphere again declines. We have focused here on CO$_2$ and H$_2$O, but the analysis described is quite general and would apply to any situation where an estimate of a condensible gases' response to addition of a non-condensible greenhouse gas is required.

Having established that we understand the fundamental behaviour of the model, we now turn to the cases where some or all of the atmosphere is allowed to evolve freely. Fig.~\ref{fig:vstrat}b) shows the cases where the temperature profile was fixed to the moist adiabat below 0.2~bar but allowed to evolve freely in the upper atmosphere. Broadly speaking, surface temperatures are similar to the $T_{strat}=200$~K case.
However, for low values of the stellar forcing $F$, $f_{H_2O}^{trap}$ is significantly lower, due to CO$_2$ cooling in the upper atmosphere. The transition to a warm, saturated stratosphere as $F$ is increased is nonlinear and rapid, due to  near-IR absorption of incoming stellar radiation by H$_2$O.

When low atmosphere inversions were permitted, the behaviour of the system was more extreme. Fig.~\ref{fig:vstrat}c) shows that in this case, $T_{surf}$ remains below 350~K for all values of $p_{CO2}$ until the solar flux is high enough for a runaway greenhouse state to occur. After this, no thermal equilibrium solutions were found for any $T_{surf}$ values between 250 and 500~K. As might be expected, the values of $f_{H_2O}^{trap}$ were correspondingly low in the pre-runaway cases. Temperature differences between the surface and warmest regions of the atmosphere reached $\sim$70~K in the most extreme scenarios (i.e., high $p_{CO_2}$, high $F$).

For the M-star case, we found broadly similar stratospheric moistening patterns as a function of $F$. The transition to a moist stratosphere tended to occur at lower $F$ values due to the decreased planetary albedo and increased high atmosphere absorption of stellar radiation,
although trapping was effective at very high $p_{CO_2}$ levels. In addition, when lower atmosphere temperature inversions were permitted, they were typically even stronger than in the G-star case.

\subsection{Sensitivity of the results to cloud assumptions}\label{subsec:clouds}

Up to this point, we have entirely neglected the effects of clouds on the atmospheric radiative budget. Clouds play a key role in the climates of Earth, past and present \citep{Goldblatt2010,Hartmann1986} and Venus \citep{Titov2007}. However, their effects are extremely hard to predict in general, due to continued uncertainty in microphysical and small-scale convective processes.
Here, to get a estimate of their effects on our main conclusions, we performed a sensitivity study involving a single H$_2$O cloud layer with 100\% coverage of the surface and an atmosphere with the same composition, temperature profile and stellar forcing as in Fig.~\ref{fig:radequi}. CO$_2$ clouds would not form in the atmospheres we are discussing because the temperatures are too high to intersect the CO$_2$ vapour-pressure curve at any altitude.

As Fig.~\ref{fig:clouds} shows, the net radiative forcing vs. the clear-sky case due to the presence of clouds is negative over a wide range of conditions. Only high clouds have a significant effect on the OLR, because at depth the longwave radiative budget is dominated by H$_2$O and CO$_2$ absorption at all wavelengths. However, high clouds are also more effective at increasing the planetary albedo. To have a warming effect, high clouds must be composed of particles that are large enough to effectively extinguish upwelling longwave radiation without significantly increasing the albedo. While this is not inconceivable, the extent of such clouds is likely to be limited due to the low residence times of larger cloud particles and lower rate of condensation in the high atmosphere.

Hence adding a more realistic representation of clouds would most likely lower surface temperatures compared to the clear-sky simulations we have discussed. This would cause even lower predictions of the H$_2$O mixing ratio at the cold trap, which is in keeping with our aim of estimating the upper limit for water loss as a function of $p_{CO_2}$. In this sense, our results are in line with previous studies, particularly \cite{Kasting1988}, who also tested the effects of clouds in their model and came to similar conclusions about their effect on climate. Some improvement in cloud modeling can be provided by 3D planetary climate simulations \citep[e.g., ][]{Wordsworth2011}, which allows the effects of the large-scale dynamics to be taken into account. However, fundamental assumptions on the nature of the cloud microphysics are still necessary in any model. Hence studies that constrain cloud effects rather than predicting them are likely by necessity to be the norm for some time to come.

\subsection{Effects of changing atmospheric nitrogen content}\label{subsec:N2}

Because of the random nature of volatile delivery to planetary atmospheres during and just after formation, it is also interesting to consider the variations in H$_2$O loss rates that occur when the N$_2$ content of a planet varies. Like H$_2$O and CO$_2$, nitrogen affects the radiative properties of the atmosphere, through collision broadening and collision-induced absorption (CIA) in the infrared and Rayleigh scattering in the visible. These effects tend to partially cancel out, with the result that the effect of doubling atmospheric N$_2$ on Earth is a small increase in surface temperature \citep{Goldblatt2009}. Hydrogen-nitrogen CIA can cause efficient warming in cases when the hydrogen content of the atmosphere is greater than a few percent \citep{Wordsworth2013b}, but we will not consider such scenarios further here.

When CO$_2$ levels are high, N$_2$ warming can be much more significant, because its
effectiveness as a Rayleigh scatterer is less than that of CO$_2$. Fig.~\ref{fig:varyN2g} shows that
a fivefold increase in the atmospheric nitrogen inventory of an Earth-like planet can cause
large surface temperature increases at high $p_{CO_2}$. Nonetheless, in terms of the cold-trap
H$_2$O mixing ratio, the thermodynamic effects of N$_2$ are most critical.
As should be clear from (\ref{eq:adia_Tlimit}), an increase in the partial pressure of the non-condensible atmospheric component means a higher surface temperature is required to keep $f_{H_2O}^{trap}$ at the same value. Fig.~\ref{fig:varyN2g} shows that
with $5\times$~PAL atmospheric N$_2$, an Earth-like planet would have a significantly
drier stratosphere despite the increase in surface temperature for $p_{CO_2}>$0.3~bar.

Conversely, if N$_2$ levels are low, upper atmosphere saturation and hence water loss can become extremely effective. In the limiting case where the N$_2$ and CO$_2$ content of the atmosphere is zero, efficient (UV energy-limited) water loss occurs at \emph{any} surface temperature. Even an ice-covered planet with surface temperatures everywhere below zero could rapidly dissociate water and lose hydrogen to space if the atmosphere was devoid of non-condensing gases. Such a scenario would likely be short-lived on an Earth-like planet, because CO$_2$ would quickly accumulate in its atmosphere due to volcanic outgassing. This would not be the case if the planet's composition was dominated by H$_2$O, as in the `super-Europa' scenarios discussed in \cite{Pierrehumbert2011b}. In this situation, however, there would be no obvious sink for O$_2$ generated by H$_2$O photolysis, so an oxygen atmosphere would presumably accumulate. This could eventually limit water loss by the cold-trap mechanism, although we note that without CO$_2$ cooling, an O$_2$-dominated upper atmosphere could reach extremely high temperatures. This issue has implications for the search for life on other planets, because oxygen is frequently considered to be a biomarker gas \citep[e.g., ][]{Selsis2002}. We leave the pursuit of this interesting problem for future research.

Finally, surface gravity affects stratospheric H$_2$O mixing ratios in predictable ways.
Increased $g$ leads to higher $p_{n,surf}$ for a given atmospheric N$_2$ inventory, reducing
$\mathcal M$ and hence stratospheric moistening for a given surface temperature.
Note, however, that the effect of this on water loss is partially mitigated by the fact that scale height
decreases with gravity, and hence diffusion-limited H$_2$O loss rates \emph{increase} [see (\ref{eq:difflimescape})].

\subsection{Water loss due to impacts}

The final modifying effect we considered was heating due to meteorite impacts. Impacts have been studied in the context of early Venus, Earth and Mars in terms of their potential to cause heating and modification of the atmosphere and surface \citep{Zahnle1988,Abramov2009,Segura2008}. Delivery of volatiles by impactors during the late stages of planet formation is also of course a major determinant of a planet's final water inventory, as we discussed in the Introduction. Here, our aim is simply to estimate whether impact heating could modify our conclusion that cold-trapping of H$_2$O strongly limits water loss for most values of $p_{CO_2}$.
First, we calculate the impactor energy required to moisten the stratosphere for a given starting composition and surface temperature. We then compare this value with the critical energy required for an impactor to cause substantial portions of the atmosphere to be directly ejected to space. In the interests of getting an upper limit on water loss, we ignore the potential for ice-rich impactors to deliver H$_2$O directly to the surface.

We assume that for an impactor of given mass and velocity, a portion $\varepsilon E_K$ of the total kinetic energy $E_K$ per unit planetary surface area will be used to directly heat the atmosphere (Fig.~\ref{fig:impactschematic}). Accounting for the sensible and latent enthalpy $E_{sens}$ and $E_{lat}$, the total energy of an atmosphere per unit surface area can be written as
\begin{eqnarray}
E_{tot} &=& E_{sens} + E_{lat} \\
        &=& \frac 1g \int^{p_{surf}}_0 \left(c_p T + q_{v}L \right)\mbox{d}p\label{eq:Eatm}
\end{eqnarray}
if we assume that the contribution of any condensed material is small and neglect the latent heat of `incondensible' components like N$_2$ and CO$_2$. Here $c_p$ is the mean constant-pressure heat capacity and $q_v = (m_v \slash \overline m)f_v$ is the mass mixing ratio of the condensable component (H$_2$O), with $\overline m$ the (local) mean molar mass of the atmosphere.

To the level of accuracy we are interested in, the initial atmospheric energy can be
approximated from (\ref{eq:Eatm}) as
\begin{equation}
E_{tot} = E_0 \sim \frac{c_{p,n}T_{surf}p_{n,surf}}{g} +
            \epsilon  L p_v(T_{surf}) / g,
\end{equation}
We now wish to calculate the threshold energy input necessary to push the atmosphere into a moist stratosphere regime. As shown previously, the transition occurs when $\mathcal M\sim 1$, and hence $E_{sens,n}\sim E_{lat}$. Given an atmospheric energy just after impact of $E_1(\mathcal M=1)$,
 the overall energy balance can be written
\begin{eqnarray}
\varepsilon E_K &=& E_1(\mathcal M=1) - E_0 \\
                &=& 2E_{lat}(T_{surf}^*) - \frac{c_{p,n}T_{surf}p_{n,surf}}{g} -  \frac{\epsilon  L p_v(T_{surf})}g.
\end{eqnarray}
Given $\mathcal M=1$, from (\ref{eq:Mdef}) we have the transcendental equation $\epsilon p_v(T_{surf}^*) L= p_{n,surf} c_{p,n} T_{surf}^*$. This can be solved for $T_{surf}^*$ for a given $p_{n,surf}$ by Newton's method, assuming 100~\% relative humidity at the surface. This then allows $\epsilon E_K$ to be calculated as a function of $p_{n,surf}$ and $T_{surf}$.

In Fig.~\ref{fig:impactimpact}, the minimum impactor radius $r_{crit}$ required to cause a transition to the moist regime is plotted vs. initial surface temperature $T_{surf}$, for three CO$_2$ partial pressures, assuming 100\% energy conversion efficiency ($\varepsilon=1$), a mean impactor density of $\rho_i = 3$~g~cm$^{-3}$, and an impact velocity equal to Earth's escape velocity. For simplicity, N$_2$ is neglected and the Clausius-Clayperon equation is used for $p_v$. Alongside this, the critical radius for erosion of a significant portion of the atmosphere $r_{erode}$ is also shown. The latter quantity can be defined as the radius required for removal of a tangent plane of the atmosphere \citep{Ahrens1993} such that
\begin{equation}
r_{erode} = \left( \frac{3}{4\pi}\frac{\rho_a}{\rho_i}H_s^2 r_p \right)^{\frac 13}
\end{equation}
where $\rho_a$ and $H_s$ are representative density and scale height values for the atmosphere and $r_p$ is the planetary radius. In Fig.~\ref{fig:impactschematic}, we use surface values for $\rho_a$ and $H_s$ to get an upper limit for $r_{erode}$.

As can be seen, the critical erosion radius is significantly smaller than the radius required to create a moist upper atmosphere except when the initial surface temperature is very close to the value at which $\mathcal M = 1$. It is therefore almost impossible for atmospheres to be forced into a moist stratosphere regime by impact heating without significant erosion also occurring. Erosion will remove a fraction of the incondensible atmospheric component of order $\frac14 H_s \slash r_p$, and has the side-effect of also making it possible for smaller subsequent impactors to cause erosion. Without any further calculation, it is therefore clear that impacts will only cause substantial water loss if they also remove significant amounts of CO$_2$ and/or N$_2$ from the planet's atmosphere.

{Interestingly, \cite{Genda2005} argued that impact erosion on planets with oceans may be quite efficient, because of the expansion of hot vaporised H$_2$O and reduced shock impedance of liquid water compared to silicate materials. Clearly, if this mechanism reduced atmospheric CO$_2$ or N$_2$ to extremely low levels post-formation on an ocean planet, water loss could then become rapid, as described in Section~\ref{subsec:N2}.
However, while ocean-enhanced erosion may have been important for removing much of Earth's primordial atmosphere when it formed, it clearly allowed substantial amounts of N$_2$ and CO$_2$ to remain, as evidenced by the significant total present-day inventories of these volatiles. For higher mass planets, it is therefore still plausible that large volatile inventories remain in the period immediately following the late stages of oligarchic growth.}

\subsection{Escape rate in moist stratosphere ($\mathcal M>>1$) limit}\label{subsec:energylim}

So far, we have only considered processes that affect water loss by modifying the saturation of H$_2$O at the cold trap. To complete the analysis, we now discuss constraints on the rate of H escape when the cold-trapping of H$_2$O in the stratosphere is no longer a limiting factor.
The first constraint we considered was the maximum possible photolysis rate of H$_2$O. We estimated this by calculating the integral
\begin{equation}
\phi_{photo}= \int_0^{\lambda_{cut}}Q_y(\lambda)F(\lambda)\mbox{d}\lambda
\end{equation}
where $F(\lambda)$ is the {net stellar flux per unit area of the planet's surface},
$Q_y$ is the quantum yield of the reaction \ce{H2O + h\nu \to H + OH} and\footnote{H$_2$O also dissociates via \ce{H2O + h\nu \to H2 + O(^1D)} and \ce{H2O + h\nu \to 2H + O(^3P)}, but the yields from these reactions are typically around two orders of magnitude lower.} $\lambda_{cut} = 196$~nm is the wavelength beyond which UV absorption by H$_2$O is negligible (see Fig.~\ref{fig:CO2_H2O_XUV}). For present-day values of the solar UV spectrum we calculated $\phi_{photo}=2\times10^{12}$~molecules~cm$^{-2}$~s$^{-1}$. This corresponds to a rapid water loss rate of 3.2 Earth oceans Gy$^{-1}$, which would be even higher under elevated XUV/UV flux conditions. We tested the dependence of this limit on the CO$_2$ mixing ratio in the upper atmosphere, but found that CO$_2$ had little shielding effect when H$_2$O was a significant atmospheric component, because the cross-section of H$_2$O is higher in the UV region (see Figs.~\ref{fig:CO2_H2O_XUV} and \ref{fig:CO2_H2O_escapebox}b).

Another limit on water loss in the saturated case can be found by considering the energy budget of the upper atmosphere. Figure~\ref{fig:CO2_H2O_escapebox} shows the results of a calculation based on the equations described in Section~\ref{subsec:escapemethod}, with the exospheric temperature at each CO$_2$ mixing ratio value found by linear interpolation to solve (\ref{eq:upperatmbox}) over a grid of values between 100 and 1000~K. In this example, we assumed a pure CO$_2$/H$_2$O upper atmosphere and synthetic solar UV spectrum appropriate to the present day, with an efficiency factor of 0.15 included in the XUV heating rate to incorporate photochemical and ionization effects \citep{Kasting1983,Chassefiere1996}.
As can be seen, the CO$_2$ mixing ratio significantly affects the escape rate, with the energetic H loss limit
slightly below the photolysis limit for low homopause $f_{CO_2}$ values, but decreasing to much lower values when CO$_2$ is a major constituent.
Nonetheless, because we neglect cooling due to H$_2$O, the escape rates at low $f_{CO_2}$ values are probably unrealistically high. Adiabatic cooling of the escaping H, which is also neglected, is also important when the escape flux is high and would tend to cause lower values of $\phi_{hydro}$ than are shown here. {Experimentation with different assumptions for $\phi_{hydro}(T_{base})$, including a conduction-free scheme that incorporates adiabatic cooling based on \cite{Pie Pierrehumbert2011BOOK}, indicated that the value of $f_{CO_2}$ at which the escape rate begins to decrease to low values is likely overestimated in our model (results not shown). Nonetheless, our calculated XUV-limited escape rate of $\sim2.2\times10^{11}$ atoms~cm$^{-2}$~s$^{-1}$ or $\sim1\times10^{30}$ atoms~s$^{-1}$ is reasonably close to values found in vertically resolved escape models that assume similar initial conditions (e.g., \cite{Erkaev2013}, Table 2). This indicates the ability of our approach to provide a basic upper limit on water loss in the presence of additional radiative forcing from UV absorption and IR emission}.

When we increased the XUV/UV flux, the H escape rate rose correspondingly. The exospheric temperature also rose somewhat,
but the efficiency of escape cooling under our isothermal wind assumption prevented it from exceeding 500~K even for a solar flux corresponding to 0.1~Ga. Under these extreme conditions, the escape rates in the model exceeded the photolysis limit even when CO$_2$ was abundant at the homopause. This result can be compared with the analysis of \cite{Kulikov2006}, who calculated exospheric temperatures in a dry Venusian atmosphere that included the effects of conduction but neglected energy removal by atmospheric escape, and estimated that rapid hydrogen escape would occur for XUV fluxes $\sim 70 \times$ present-day.

Our simple model allows several basic conclusions to be drawn regarding water loss in the moist stratosphere limit. First in agreement with \cite{Kulikov2006}, we find that for planets receiving stellar fluxes that place them close to or over the runaway limit, such as early Venus, H removal could probably have been rapid even if CO$_2$ was abundant in the atmosphere. However, planets with CO$_2$-rich atmospheres around G-stars that receive a similar stellar flux to Earth can only experience significant UV-powered water loss early in their system's lifetime. Around M-stars, XUV levels are elevated for much longer and the stellar luminosity is essentially unvarying with time, so more escape may occur if
water vapour is abundant in the high atmosphere. The differences between G- and M-star cases and implications for terrestrial exoplanets in general are discussed further in the following section.

\section{Discussion}\label{sec:discuss}

\subsection{H$_2$O loss rates vs. atmospheric CO$_2$ pressure}

To get an integrated view of water loss rates under a wide range of conditions, we used (\ref{eq:difflimescape}) in combination with (\ref{eq:FYS}) and the calculations discussed in the previous sections. {In cases where the stellar flux was high enough to cause a runaway greenhouse, upper atmospheric $f_{H_2O}$ and $f_{CO_2}$ were calculated assuming a well mixed atmosphere and a total H$_2$O inventory of 1 Earth ocean for simplicity.} The results in terms of Earth oceans per Gy are displayed in Figure~\ref{fig:escape_total} as a function of time/orbital distance and atmospheric CO$_2$. As can be seen, for the G-star case, water loss is diffusion-limited (and low) until late in the Sun's evolutionary history, when surface temperatures increase sufficiently to allow a moist stratosphere at $p_{CO_2}$ values between 0.1 and 1~bar.

The fact that XUV and FUV fluxes decrease with time but \emph{total} solar luminosity increases with time makes water loss from Earth-like planets around G-stars particularly hard to achieve. The faint young Sun effect causes strong limits on $f_{H_2O}$ at the cold trap early on for all values of $p_{CO_2}$. However, by the time total luminosity has increased enough to allow an H$_2$O-rich stratosphere at moderate $p_{CO_2}$ values, the planet is near the runaway greenhouse transition, and XUV and FUV fluxes have declined enough to make energy limitations important.  For Earth, this suggests that factors such as a weaker magnetic field in the early Archean \citep{Tarduno2010} are unlikely to have led to significant water loss compared to the present-day ocean volume. Hence despite the advances in radiative transfer modelling over the last few decades, the conclusions of \cite{Kasting1986} remain essentially valid.

Around M-stars, the lack of temporal variation in total solar luminosity means water loss is most effective close to the inner
edge of the habitable zone. However, the high and unpredictable variability in the XUV/UV flux is also important.
In Fig.~\ref{fig:escape_total}b), the escape rates are plotted assuming a synthetic UV spectrum appropriate to GJ~436, which
is a relatively quiet M3 star. Fig.~\ref{fig:escape_total}c) shows results for the same case, except with the 122~nm Lyman~$\alpha$ emission line in the incident stellar spectrum scaled to the value for AU~Mic, a young and active M1 star \citep{Linsky2013}. As can be seen, Lyman~$\alpha$ variability can make a significant difference to water loss rates around M-stars both beyond and inside the runaway greenhouse threshold. Nonetheless, because of the cold trap constraints discussed in Section~\ref{subsec:dependence}, high H$_2$O loss is never achieved for planets
receiving total fluxes much less than that of Earth [approx. $d>0.17$~AU in Fig.~\ref{fig:escape_total}c)].  The only effective way to enhance H$_2$O photolysis rates in these cases would appear to be via decreases in the total atmospheric non-condensible gas content.

Taken together, these results suggest that rocky exoplanets in the habitable zone may retain even a limited water inventory if they form with little H$_2$O, which is clearly a positive outcome from a habitability standpoint. Conversely, most planets that form with much more H$_2$O than Earth are unlikely to lose it via escape. Ocean planets may therefore be relatively common in general, which, as we discuss in the next section, has important implications for the search for exoplanet biosignatures.

For Venus, it might appear obvious from our calculations that the planet has always been in a runaway state. Indeed, our clear-sky calculations suggest Earth itself receives close to the limiting runaway flux at present, in agreement with the recent results of \cite{Goldblatt2013} and \cite{Kopparapu2013}. When the Solar System formed, Venus received a solar flux $\sim1.4$ times that of Earth today, apparently placing it well inside the runaway limit. However, our calculations neglect cloud radiative forcing and spatial variations in relative humidity, both of which can have a major effect on the runaway threshold. Using the present-day atmospheric CO$_2$ inventory (92~bars) and a solar flux $F=0.72F_0$ appropriate to 4.4~Ga, for early Venus we calculate that a negative radiative forcing of around 70~W~m$^{-2}$ is needed to reach equilibrium surface temperatures of $\sim$320~K, at which point diffusion limits on H$_2$O escape are important. Hence while it is possible that clouds could have limited water loss from an early CO$_2$-dominated atmosphere, until their effects are understood in detail the argument that Venus lost its water early via rapid hydrodynamic escape \citep{Gillmann2009} remains entirely plausible.

\subsection{Climate and habitability of waterworlds}

As described in Section~\ref{sec:intro}, a planet with no subaerial land by definition will no longer experience land silicate weathering.
For waterworlds, a large fraction of the total CO$_2$ inventory would then be expected to reside in the atmosphere and ocean, unless seafloor weathering were extremely effective\footnote{In cases where surface liquid H$_2$O is a significant fraction of the planetary mass ($>$20-30 Earth oceans), volatile outgassing can become suppressed by overburden pressure \citep{Kite2009,Elkins2011}, and interior mechanisms involving clathrate hydrate formation may become important \citep{Levi2013}. It is difficult to predict how the atmospheric CO$_2$ inventory would behave in such circumstances without further coupled atmosphere-interior modelling. However, some of the arguments in the Appendix relating to atmosphere/ocean volatile partitioning would still be applicable in these situations.}.
Partitioning of CO$_2$ between the atmosphere and ocean depends on carbonate ion chemistry and hence on the ocean pH, but in the absence of major buffering effects from other species\footnote{Ammonia is soluble and weakly basic in water, and hence could conceivably buffer ocean pH if it was present in large enough quantities, but it is efficiently converted to N$_2$ by photolysis in non-reducing atmospheres.}, a large fraction of the total surface CO$_2$ inventory would still remain in the atmosphere for a planet with 10 times Earth's ocean amount (see the Appendix for details). We have just shown that water loss rates in CO$_2$-rich atmospheres will be low for a wide range of conditions, so waterworlds could plausibly remain stable throughout their history. In the context of future searches for biosignatures on other planets [e.g., \cite{Kaltenegger2013}], therefore, it is interesting to consider the potential differences in habitability that are likely when no subaerial land is present.

Aside from the presence of liquid water, the first major consideration for the survival of life is surface temperature. If waterworlds do tend
to have high atmospheric CO$_2$ inventories, those receiving an Earth-like stellar flux would have surface temperatures in the 350-450~K range. The survival range for life on Earth is around 250 to 400~K \citep{Kashefi2003}, so a waterworld could perhaps still remain marginally habitable by this criterion unless other warming mechanisms were also present (see Figure~\ref{fig:vstrat}).

Other constraints may come from the potential for life to emerge in the first place. It has been argued that life on Earth originated in shallow ocean or coastal regions, with evaporation cycles playing a key role in the development of a `primordial soup' \citep{Bada2004}. Such a scenario would clearly be impossible on a planet with no exposed rock at the surface. Another leading hypothesis for the origin of life on Earth posits that it occurred in hydrothermal vents (specifically, in alkaline vents similar to the Lost City region in the mid-Atlantic) \citep{Russell1994,Kelley2005}. However, even this mechanism could become problematic if the ocean volume is so large that pressures at the seafloor are high enough to inhibit outgassing \citep[e.g., ][]{Lammer2009}.

Finally, besides liquid water and an equable temperature range, all life on Earth requires certain essential nutrients (the so-called `CHNOPS' elements plus a variety of metals). In the present-day oceans, net primary production is believed to be limited ultimately by the availability of phosphorous in particular, which is delivered primarily by weathering of exposed rock on the surface \citep{Filippelli2008}. Photosynthetic life in the ocean is restricted to the surface euphotic layer, but in the absence of a land source, elements like phosphorous, iron and sulphur could only be supplied there from the ocean floor, at rates that are typically 2-3 orders of magnitude smaller than comparable supply from the continents \citep{Kharecha2005}. An Earth-like biosphere on a waterworld would therefore have a net primary productivity that was several orders of magnitude lower than that of Earth today (see Figure~\ref{fig:oceanschematic}). Given the strong selection pressures that would be present in such a nutrient-poor environment, it is conceivable that organisms dependent only on elements accessible from the atmosphere could develop.
Nonetheless, these general considerations hint at some of the differences we should expect between land planets and ocean planets, as well as the subtlety of the relationship between water and habitability in general. Rather than simply extrapolating Earth-like atmospheric conditions and biospheric productivity, future biosignature studies should aim to investigate these issues in more detail.

\subsection{Future work}\label{sec:future}

There are a number of potential future research directions from this study. First, our results clearly indicate the need for a greater understanding of how the crust and mantle of Earth-like planets with high H$_2$O inventories evolve. Here, we have focused on the atmospheric component of the problem, but large uncertainties still remain regarding the exchange of CO$_2$ and H$_2$O between a planet's mantle and surface. For CO$_2$, the high uncertainty in the physics and chemistry of seafloor weathering currently limits our ability to extrapolate Earth's climate evolution to more general cases. This is a problem that would benefit greatly from more detailed observational and experimental constraints. For H$_2$O, partitioning between the surface and mantle is also still poorly understood \citep{Hirschmann2006}, although it has been hypothesized that if Earth's ocean volume was \emph{lower}, it would increase to the present value due to a feedback involving the ridge axis hydrothermal circulation \citep{Kasting1992}. If an (as yet unidentified) negative crust-mantle feedback also operates in the other direction, our conclusions regarding the potential abundance of ocean planets could require revision.

Regarding climate modeling, an obvious extension of this work is to examine the role of clouds and relative humidity variations in detail using a 3D climate model. For tidally locked planets around M-stars, in particular, the differences in the 3D case could be significant, because the nonlinear dependence of $f_{H_2O}^{trap}$ on stellar forcing means that the planet's dayside stratosphere could be much more humid than a global mean calculation would suggest. We plan to assess the differences caused by the transition to 3D in future work. GCMs are also able to tackle cloud effects more accurately in principle, although as we have mentioned, uncertainties in sub-gridscale processes and cloud microphysics are not removed by 3D modeling. Selected numerical experiments using cloud-resolving models, perhaps combined with direct laboratory experiments on cloud microphysics under a range of non-Earth-like conditions, would be a valuable way to gain insight in future. Nonetheless, despite the uncertainties, the fact that clouds cool over most conditions relative to the clear-sky case means that they are unlikely to affect the robustness of our general conclusions here.

Observationally, we are still some way from being able to characterize low mass exoplanets of the type we have discussed, although the state of the art is advancing rapidly \citep{Bean2010,Croll2011,Kreidberg2013}. Both JWST and ESA's planned EChO mission will be able to perform spectroscopic analysis of the atmospheres of nearby transiting super-Earths, which at minimum will allow the major optically active species in their atmospheres to be identified. However, to distinguish planets with volatile-rich atmospheres and high surface temperatures from more Earth-like cases, characterization of absorber abundances and surface pressures will be a key challenge. This can be done by transmission spectroscopy, in principle, as long as the planet's atmosphere is clear enough in the visible at short wavelengths to allow identification of the spectral Rayleigh scattering slope \citep{Benneke2012}. Another promising approach that is valid for non-transiting planets is spectral phase curve analysis \citep{Selsis2011}, although the demands on instrumental sensitivity with this method are stringent. In the long term, detailed observational tests of planetary water loss theories will be best achieved via revival of NASA and ESA's TPF/Darwin exoplanet characterization missions.

\section{Acknowledgments}
Photodissociation cross-section and quantum yield data and the solar spectrum in the UV were kindly provided by E. H\'ebrard at the Universit\'e de Bordeaux. The code used to compute the moist adiabat was partly based on routines originally provided by E.~Marcq. For the M-star UV spectrum, we acknowledge use of the MUSCLES database. R. W. thanks Ty Robinson for enlightening intercomparisons with the SMART radiative code, and F. Ciesla, K. France, J. Linsky, R. Heller, D. Abbot and N. Cohen for discussions.

\appendix

\section{Ocean / atmosphere partitioning of CO$_2$ on water-rich planets}\label{apx:oceanpH}

To calculate the fraction of \ce{CO_2} stored in the ocean for a given atmospheric partial pressure, we calculated the chemical equilibria of the \ce{CO_2}-carbonate-bicarbonate system, assuming contact with an infinite calcium carbonate reservoir following the methodology described in \cite{Pierrehumbert2011BOOK}. Chemical equations
\begin{equation}
\ce{CO_2(aq) + H_2O} \rightleftharpoons \ce{HCO_3^- + H^+}\label{eq:chem1}
\end{equation}
\begin{equation}
\ce{HCO_3^-} \rightleftharpoons \ce{CO_3^{2-} + H^+}\label{eq:chem2}
\end{equation}
and
\begin{equation}
\ce{CaCO_3(s)} \rightleftharpoons \ce{Ca^{2+} + CO_3^{2-}}\label{eq:chem3}
\end{equation}
were solved for a given pH by Newtonian iteration using the corresponding charge balance equation. Ocean \ce{CO_{2(aq)}} was related to atmospheric $p_{CO_2}$ via Henry's Law. Equilibrium and constants and their temperature dependencies were calculated from data in Tables 8.1 and 8.2 of \cite{Pierrehumbert2011BOOK}, {while for the Henry's Law coefficient, data from \cite{Carroll1991} was used}. Finally, the ratio of atmospheric to ocean carbon content was calculated as
\begin{equation}
\mathcal R_C = \frac{K_H N_{CO_2(g)}}{p_{CO_2} N_{H_2O(l)}}\frac{1}{1+K_1\ce{[H^+]}^{-1} + K_1K_2\ce{[H^+]}^{-2}}
\end{equation}
with $K_1(T)$ and $K_2(T)$ the equilibrium constants of (\ref{eq:chem1}) and (\ref{eq:chem2}), respectively, $K_H(T)$ Henry's constant for \ce{CO_2}, and $N_{H_2O(l)}$ and $N_{CO_2(g)}$ the total number of moles of \ce{H_2O} in the ocean and \ce{CO_2} in the atmosphere, respectively. The latter quantity was calculated as a function of $p_{CO_2}$ and $T_{surf}$ using the atmospheric code described in the main text. Figure~\ref{fig:ocean_C_content} shows $\mathcal R_C$ as a function of $p_{CO_2}$ for various ocean temperatures, for a hypothetical super-Earth exoplanet with $g=15.0$~m~s$^{-2}$, total ocean amount 10~$\times$ that of Earth and radius $r_P = 1.3 r_E$.

As can be seen, $\mathcal R_C$ increases rapidly with $p_{CO_2}$ in all cases, increasing to over 0.1 for $p_{CO_2}>0.25$~bar at $T_{surf}=300$~K despite the increased ocean volume. $\mathcal R_C$ also significantly increases with temperature for all $p_{CO_2}$ values. This is primarily because $K_H$ (and hence CO$_2$ solubility) decreases with temperature, limiting the total amount of inorganic carbon the ocean can hold.  This effect may be important for ocean planet climates in general: given the dependence of ocean temperatures on atmospheric \ce{CO_2} via the greenhouse effect, this should lead to a \emph{positive} feedback on ocean planets between $p_{CO_2}$ and $T_{surf}$, which clearly will have a destabilizing effect. {Because the solubility of many significant greenhouse gases decreases with temperature in water over wide ranges}, similar positive feedbacks involving other gases could also be significant on ocean planets.

\bibliography{allrefs}

\begin{thebibliography}{109}
\providecommand{\natexlab}[1]{#1}
\providecommand{\url}[1]{\texttt{#1}}
\expandafter\ifx\csname urlstyle\endcsname\relax
  \providecommand{\doi}[1]{doi: #1}\else
  \providecommand{\doi}{doi: \begingroup \urlstyle{rm}\Url}\fi

\bibitem[Abbot et~al.(2012)Abbot, Cowan, and Ciesla]{Abbot2012}
D.~S. Abbot, N.~B. Cowan, and F.~J. Ciesla.
\newblock {Indication of insensitivity of planetary weathering behavior and
  habitable zone to surface land fraction}.
\newblock \emph{The Astrophysical Journal}, 178:\penalty0 756, 2012.
\newblock \doi{10.1088/0004-637X/756/2/178}.

\bibitem[Abe et~al.(2011)Abe, Abe-Ouchi, Sleep, and Zahnle]{Abe2011}
Yutaka Abe, Ayako Abe-Ouchi, Norman~H Sleep, and Kevin~J Zahnle.
\newblock Habitable zone limits for dry planets.
\newblock \emph{Astrobiology}, 11\penalty0 (5):\penalty0 443--460, 2011.

\bibitem[{Abramov} and {Mojzsis}(2009)]{Abramov2009}
O.~{Abramov} and S.~J. {Mojzsis}.
\newblock {Microbial habitability of the Hadean Earth during the late heavy
  bombardment}.
\newblock \emph{Nature}, 459:\penalty0 419--422, May 2009.
\newblock \doi{10.1038/nature08015}.

\bibitem[Ahrens(1993)]{Ahrens1993}
Thomas~J Ahrens.
\newblock Impact erosion of terrestrial planetary atmospheres.
\newblock \emph{Annual Review of Earth and Planetary Sciences}, 21:\penalty0
  525--555, 1993.

\bibitem[Bada(2004)]{Bada2004}
Jeffrey~L Bada.
\newblock How life began on earth: a status report.
\newblock \emph{Earth and Planetary Science Letters}, 226\penalty0
  (1):\penalty0 1--15, 2004.

\bibitem[{Baranov} et~al.(2004){Baranov}, {Lafferty}, and
  {Fraser}]{Baranov2004}
Y.~I. {Baranov}, W.~J. {Lafferty}, and G.~T. {Fraser}.
\newblock {Infrared spectrum of the continuum and dimer absorption in the
  vicinity of the O2 vibrational fundamental in O2/CO2 mixtures}.
\newblock \emph{J. Mol. Spectrosc.}, 228:\penalty0 432--440, December 2004.
\newblock \doi{10.1016/j.jms.2004.04.010}.

\bibitem[{Bean} et~al.(2010){Bean}, {Kempton}, and {Homeier}]{Bean2010}
J.~L. {Bean}, {E.~M.-R.} {Kempton}, and D.~{Homeier}.
\newblock {A ground-based transmission spectrum of the super-Earth exoplanet GJ
  1214b}.
\newblock \emph{Nature}, 468:\penalty0 669--672, December 2010.
\newblock \doi{10.1038/nature09596}.

\bibitem[Benneke and Seager(2012)]{Benneke2012}
Bjoern Benneke and Sara Seager.
\newblock Atmospheric retrieval for super-earths: Uniquely constraining the
  atmospheric composition with transmission spectroscopy.
\newblock \emph{The Astrophysical Journal}, 753\penalty0 (2):\penalty0 100,
  2012.

\bibitem[Brown et~al.(2007)Brown, Humphrey, and Gamache]{Brown2007}
LR~Brown, CM~Humphrey, and RR~Gamache.
\newblock {CO$_2$-broadened water in the pure rotation and $\nu_2$ fundamental
  regions}.
\newblock \emph{Journal of Molecular Spectroscopy}, 246\penalty0 (1):\penalty0
  1--21, 2007.

\bibitem[Caldeira(1995)]{Caldeira1995}
Ken Caldeira.
\newblock Long-term control of atmospheric carbon dioxide; low-temperature
  seafloor alteration or terrestrial silicate-rock weathering?
\newblock \emph{American Journal of Science}, 295\penalty0 (9):\penalty0
  1077--1114, 1995.

\bibitem[Carroll et~al.(1991)Carroll, Slupsky, and Mather]{Carroll1991}
John~J Carroll, John~D Slupsky, and Alan~E Mather.
\newblock The solubility of carbon dioxide in water at low pressure.
\newblock \emph{J. Phys. Chem. Ref. Data}, 20\penalty0 (6):\penalty0
  1201--1209, 1991.

\bibitem[Chan et~al.(1993{\natexlab{a}})Chan, Cooper, and Brion]{Chan1993}
WF~Chan, G~Cooper, and CE~Brion.
\newblock The electronic spectrum of water in the discrete and continuum
  regions. absolute optical oscillator strengths for photoabsorption (6--200
  ev).
\newblock \emph{Chemical physics}, 178\penalty0 (1):\penalty0 387--400,
  1993{\natexlab{a}}.

\bibitem[Chan et~al.(1993{\natexlab{b}})Chan, Cooper, Sodhi, and
  Brion]{Chan1993b}
WF~Chan, G~Cooper, RNS Sodhi, and CE~Brion.
\newblock Absolute optical oscillator strengths for discrete and continuum
  photoabsorption of molecular nitrogen (11--200 ev).
\newblock \emph{Chemical physics}, 170\penalty0 (1):\penalty0 81--97,
  1993{\natexlab{b}}.

\bibitem[{Chassefi{\`e}re}(1996)]{Chassefiere1996}
E.~{Chassefi{\`e}re}.
\newblock {Hydrodynamic escape of hydrogen from a hot water-rich atmosphere:
  The case of Venus}.
\newblock \emph{Journal of Geophysical Research}, 101:\penalty0 26039--26056,
  November 1996.
\newblock \doi{10.1029/96JE01951}.

\bibitem[{Chassefi{\`e}re} et~al.(2012){Chassefi{\`e}re}, {Wieler}, {Marty},
  and {Leblanc}]{Chassefiere2012}
E.~{Chassefi{\`e}re}, R.~{Wieler}, B.~{Marty}, and F.~{Leblanc}.
\newblock {The evolution of Venus: Present state of knowledge and future
  exploration}.
\newblock \emph{Planetary and Space Science}, 63:\penalty0 15--23, April 2012.
\newblock \doi{10.1016/j.pss.2011.04.007}.

\bibitem[Clough et~al.(1989)Clough, Kneizys, and Davies]{Clough1989}
S.A. Clough, F.X. Kneizys, and R.W. Davies.
\newblock Line shape and the water vapor continuum.
\newblock \emph{Atmospheric Research}, 23\penalty0 (3-4):\penalty0 229 -- 241,
  1989.
\newblock ISSN 0169-8095.

\bibitem[Cranmer(2004)]{Cranmer2004}
Steven~R Cranmer.
\newblock {New views of the solar wind with the Lambert W function}.
\newblock \emph{American Journal of Physics}, 72:\penalty0 1397, 2004.

\bibitem[Croll et~al.(2011)Croll, Albert, Jayawardhana, Kempton, Fortney,
  Murray, and Neilson]{Croll2011}
Bryce Croll, Loic Albert, Ray Jayawardhana, Eliza Miller-Ricci Kempton,
  Jonathan~J Fortney, Norman Murray, and Hilding Neilson.
\newblock Broadband transmission spectroscopy of the super-earth gj 1214b
  suggests a low mean molecular weight atmosphere.
\newblock \emph{The Astrophysical Journal}, 736\penalty0 (2):\penalty0 78,
  2011.

\bibitem[{de Bergh} et~al.(1991){de Bergh}, {Bezard}, {Owen}, {Crisp},
  {Maillard}, and {Lutz}]{deBergh1991}
C.~{de Bergh}, B.~{Bezard}, T.~{Owen}, D.~{Crisp}, J.-P. {Maillard}, and B.~L.
  {Lutz}.
\newblock {Deuterium on Venus - Observations from Earth}.
\newblock \emph{Science}, 251:\penalty0 547--549, February 1991.
\newblock \doi{10.1126/science.251.4993.547}.

\bibitem[{Edson} et~al.(2012){Edson}, {Kasting}, {Pollard}, {Lee}, and
  {Bannon}]{Edson2012}
A.~R. {Edson}, J.~F. {Kasting}, D.~{Pollard}, S.~{Lee}, and P.~R. {Bannon}.
\newblock {The Carbonate-Silicate Cycle and CO2/Climate Feedbacks on Tidally
  Locked Terrestrial Planets}.
\newblock \emph{Astrobiology}, 12:\penalty0 562--571, June 2012.
\newblock \doi{10.1089/ast.2011.0762}.

\bibitem[Elkins-Tanton(2011)]{Elkins2011}
Linda~T Elkins-Tanton.
\newblock Formation of early water oceans on rocky planets.
\newblock \emph{Astrophysics and Space Science}, 332\penalty0 (2):\penalty0
  359--364, 2011.

\bibitem[Erkaev et~al.(2013)Erkaev, Lammer, Odert, Kulikov, Kislyakova,
  Khodachenko, G{\"u}del, Hanslmeier, and Biernat]{Erkaev2013}
Nikolai~V Erkaev, Helmut Lammer, Petra Odert, Yu~N Kulikov, Kristina~G
  Kislyakova, Maxim~L Khodachenko, Manuel G{\"u}del, Arnold Hanslmeier, and
  Helfried Biernat.
\newblock Xuv exposed non-hydrostatic hydrogen-rich upper atmospheres of
  terrestrial planets. part i: Atmospheric expansion and thermal escape.
\newblock \emph{arXiv preprint arXiv:1212.4982}, 2013.

\bibitem[Filippelli(2008)]{Filippelli2008}
Gabriel~M Filippelli.
\newblock The global phosphorus cycle: Past, present, and future.
\newblock \emph{Elements}, 4\penalty0 (2):\penalty0 89--95, 2008.

\bibitem[Fillion et~al.(2004)Fillion, Ruiz, Yang, Castillejo, Rostas, and
  Lemaire]{Fillion2004}
J-H Fillion, J~Ruiz, X-F Yang, M~Castillejo, F~Rostas, and J-L Lemaire.
\newblock High resolution photoabsorption and photofragment fluorescence
  spectroscopy of water between 10.9 and 12 ev.
\newblock \emph{The Journal of chemical physics}, 120:\penalty0 6531, 2004.

\bibitem[France et~al.(2013)France, Froning, Linsky, Roberge, Stocke, Tian,
  Bushinsky, D{\'e}sert, Mauas, Vieytes, et~al.]{France2013}
Kevin France, Cynthia~S Froning, Jeffrey~L Linsky, Aki Roberge, John~T Stocke,
  Feng Tian, Rachel Bushinsky, Jean-Michel D{\'e}sert, Pablo Mauas, Mariela
  Vieytes, et~al.
\newblock The ultraviolet radiation environment around m dwarf exoplanet host
  stars.
\newblock \emph{The Astrophysical Journal}, 763\penalty0 (2):\penalty0 149,
  2013.

\bibitem[Fu et~al.(2010)Fu, O'Connell, and Sasselov]{Fu2010}
Roger Fu, Richard~J O'Connell, and Dimitar~D Sasselov.
\newblock The interior dynamics of water planets.
\newblock \emph{The Astrophysical Journal}, 708\penalty0 (2):\penalty0 1326,
  2010.

\bibitem[Genda and Abe(2005)]{Genda2005}
Hidenori Genda and Yutaka Abe.
\newblock Enhanced atmospheric loss on protoplanets at the giant impact phase
  in the presence of oceans.
\newblock \emph{Nature}, 433\penalty0 (7028):\penalty0 842--844, 2005.

\bibitem[{Gillmann} et~al.(2009){Gillmann}, {Chassefi{\`e}re}, and
  {Lognonn{\'e}}]{Gillmann2009}
C.~{Gillmann}, E.~{Chassefi{\`e}re}, and P.~{Lognonn{\'e}}.
\newblock {A consistent picture of early hydrodynamic escape of Venus
  atmosphere explaining present Ne and Ar isotopic ratios and low oxygen
  atmospheric content}.
\newblock \emph{Earth and Planetary Science Letters}, 286:\penalty0 503--513,
  September 2009.
\newblock \doi{10.1016/j.epsl.2009.07.016}.

\bibitem[Goldblatt and Zahnle(2010)]{Goldblatt2010}
C~Goldblatt and KJ~Zahnle.
\newblock Clouds and the faint young sun paradox.
\newblock \emph{Climate of the Past Discussions}, 6:\penalty0 1163--1207, 2010.

\bibitem[{Goldblatt} et~al.(2009){Goldblatt}, {Claire}, {Lenton}, {Matthews},
  {Watson}, and {Zahnle}]{Goldblatt2009}
C.~{Goldblatt}, M.~W. {Claire}, T.~M. {Lenton}, A.~J. {Matthews}, A.~J.
  {Watson}, and K.~J. {Zahnle}.
\newblock {Nitrogen-enhanced greenhouse warming on early Earth}.
\newblock \emph{Nature Geoscience}, 2:\penalty0 891--896, December 2009.
\newblock \doi{10.1038/ngeo692}.

\bibitem[Goldblatt et~al.(2013)Goldblatt, Robinson, Zahnle, and
  Crisp]{Goldblatt2013}
Colin Goldblatt, Tyler~D Robinson, Kevin~J Zahnle, and David Crisp.
\newblock Low simulated radiation limit for runaway greenhouse climates.
\newblock \emph{Nature Geoscience}, 2013.

\bibitem[{Gough}(1981)]{Gough1981}
D.~O. {Gough}.
\newblock {Solar interior structure and luminosity variations}.
\newblock \emph{Solar Physics}, 74:\penalty0 21--34, November 1981.

\bibitem[{Gruszka} and {Borysow}(1997)]{Gruszka1997}
M.~{Gruszka} and A.~{Borysow}.
\newblock {Roto-Translational Collision-Induced Absorption of CO2 for the
  Atmosphere of Venus at Frequencies from 0 to 250 cm\^{}-1, at Temperatures
  from 200 to 800 K}.
\newblock \emph{Icarus}, 129:\penalty0 172--177, September 1997.
\newblock \doi{10.1006/icar.1997.5773}.

\bibitem[Haar et~al.(1984)Haar, Gallagher, Kell, and (U.S.)]{Haar1984}
L.~Haar, J.~Gallagher, G.~Kell, and National Standard Reference Data~System
  (U.S.).
\newblock \emph{{NBS/NRC Steam Tables: Thermodynamic and Transport Properties
  and Computer Programs for Vapor and Liquid States of Water in SI Units}}.
\newblock Hemisphere, Washington, D.C., 1984.

\bibitem[Hartmann et~al.(1986)Hartmann, Ramanathan, Berroir, and
  Hunt]{Hartmann1986}
DL~Hartmann, V~Ramanathan, A~Berroir, and GE~Hunt.
\newblock Earth radiation budget data and climate research.
\newblock \emph{Reviews of Geophysics}, 24\penalty0 (2):\penalty0 439--468,
  1986.

\bibitem[Hirschmann(2006)]{Hirschmann2006}
Marc~M Hirschmann.
\newblock {Water, melting, and the deep Earth H2O cycle}.
\newblock \emph{Annu. Rev. Earth Planet. Sci.}, 34:\penalty0 629--653, 2006.

\bibitem[Huebner et~al.(1992)Huebner, Keady, and Lyon]{Huebner1992}
Walter~F Huebner, John~Joseph Keady, and SP~Lyon.
\newblock Solar photo rates for planetary atmospheres and atmospheric
  pollutants.
\newblock \emph{Astrophysics and Space Science}, 195\penalty0 (1):\penalty0
  1--294, 1992.

\bibitem[{Ingersoll}(1969)]{Ingersoll1969}
A.~P. {Ingersoll}.
\newblock {The Runaway Greenhouse: A History of Water on Venus.}
\newblock \emph{Journal of Atmospheric Sciences}, 26:\penalty0 1191--1198,
  November 1969.
\newblock \doi{10.1175/1520-0469(1969)026<1191:TRGAHO>2.0.CO;2}.

\bibitem[Kaltenegger et~al.(2013)Kaltenegger, Sasselov, and
  Rugheimer]{Kaltenegger2013}
L~Kaltenegger, D~Sasselov, and S~Rugheimer.
\newblock Water planets in the habitable zone: Atmospheric chemistry,
  observable features, and the case of kepler-62e and-62f.
\newblock \emph{arXiv preprint arXiv:1304.5058}, 2013.

\bibitem[Kashefi and Lovley(2003)]{Kashefi2003}
Kazem Kashefi and Derek~R Lovley.
\newblock Extending the upper temperature limit for life.
\newblock \emph{Science}, 301\penalty0 (5635):\penalty0 934--934, 2003.

\bibitem[{Kasting}(1988)]{Kasting1988}
J.~F. {Kasting}.
\newblock {Runaway and moist greenhouse atmospheres and the evolution of earth
  and Venus}.
\newblock \emph{Icarus}, 74:\penalty0 472--494, 1988.
\newblock \doi{10.1016/0019-1035(88)90116-9}.

\bibitem[{Kasting} and {Ackerman}(1986)]{Kasting1986}
J.~F. {Kasting} and T.~P. {Ackerman}.
\newblock {Climatic consequences of very high carbon dioxide levels in the
  earth's early atmosphere}.
\newblock \emph{Science}, 234:\penalty0 1383--1385, 1986.
\newblock \doi{10.1126/science.11539665}.

\bibitem[{Kasting} and {Holm}(1992)]{Kasting1992}
J.~F. {Kasting} and N.~G. {Holm}.
\newblock {What determines the volume of the oceans?}
\newblock \emph{Earth and Planetary Science Letters}, 109:\penalty0 507--515,
  April 1992.
\newblock \doi{10.1016/0012-821X(92)90110-H}.

\bibitem[{Kasting} and {Pollack}(1983)]{Kasting1983}
J.~F. {Kasting} and J.~B. {Pollack}.
\newblock {Loss of water from Venus. I - Hydrodynamic escape of hydrogen}.
\newblock \emph{Icarus}, 53:\penalty0 479--508, March 1983.
\newblock \doi{10.1016/0019-1035(83)90212-9}.

\bibitem[{Kasting} et~al.(1993){Kasting}, {Whitmire}, and
  {Reynolds}]{Kasting1993}
J.~F. {Kasting}, D.~P. {Whitmire}, and R.~T. {Reynolds}.
\newblock {Habitable Zones around Main Sequence Stars}.
\newblock \emph{Icarus}, 101:\penalty0 108--128, January 1993.
\newblock \doi{10.1006/icar.1993.1010}.

\bibitem[{Kelley} et~al.(2005){Kelley}, {Karson}, {Fr{\"u}h-Green}, {Yoerger},
  {Shank}, {Butterfield}, {Hayes}, {Schrenk}, {Olson}, {Proskurowski},
  {Jakuba}, {Bradley}, {Larson}, {Ludwig}, {Glickson}, {Buckman}, {Bradley},
  {Brazelton}, {Roe}, {Elend}, {Delacour}, {Bernasconi}, {Lilley}, {Baross},
  {Summons}, and {Sylva}]{Kelley2005}
D.~S. {Kelley}, J.~A. {Karson}, G.~L. {Fr{\"u}h-Green}, D.~R. {Yoerger}, T.~M.
  {Shank}, D.~A. {Butterfield}, J.~M. {Hayes}, M.~O. {Schrenk}, E.~J. {Olson},
  G.~{Proskurowski}, M.~{Jakuba}, A.~{Bradley}, B.~{Larson}, K.~{Ludwig},
  D.~{Glickson}, K.~{Buckman}, A.~S. {Bradley}, W.~J. {Brazelton}, K.~{Roe},
  M.~J. {Elend}, A.~{Delacour}, S.~M. {Bernasconi}, M.~D. {Lilley}, J.~A.
  {Baross}, R.~E. {Summons}, and S.~P. {Sylva}.
\newblock {A Serpentinite-Hosted Ecosystem: The Lost City Hydrothermal Field}.
\newblock \emph{Science}, 307:\penalty0 1428--1434, March 2005.
\newblock \doi{10.1126/science.1102556}.

\bibitem[Kharecha et~al.(2005)Kharecha, Kasting, and Siefert]{Kharecha2005}
P.~Kharecha, J.~Kasting, and J.~Siefert.
\newblock {A coupled atmosphere-ecosystem model of the early Archean Earth}.
\newblock \emph{Geobiology}, 3:\penalty0 53--76, 2005.

\bibitem[Khodachenko et~al.(2007)Khodachenko, Ribas, Lammer, Grie{\ss}meier,
  Leitner, Selsis, Eiroa, Hanslmeier, Biernat, Farrugia,
  et~al.]{Khodachenko2007}
Maxim~L Khodachenko, Ignasi Ribas, Helmut Lammer, Jean-Mathias Grie{\ss}meier,
  Martin Leitner, Franck Selsis, Carlos Eiroa, Arnold Hanslmeier, Helfried~K
  Biernat, Charles~J Farrugia, et~al.
\newblock Coronal mass ejection (cme) activity of low mass m stars as an
  important factor for the habitability of terrestrial exoplanets. i. cme
  impact on expected magnetospheres of earth-like exoplanets in close-in
  habitable zones.
\newblock \emph{Astrobiology}, 7\penalty0 (1):\penalty0 167--184, 2007.

\bibitem[{Kite} et~al.(2011){Kite}, {Gaidos}, and {Manga}]{Kite2011}
E.~S. {Kite}, E.~{Gaidos}, and M.~{Manga}.
\newblock {Climate Instability on Tidally Locked Exoplanets}.
\newblock \emph{The Astrophysical Journal}, 743:\penalty0 41, December 2011.
\newblock \doi{10.1088/0004-637X/743/1/41}.

\bibitem[Kite et~al.(2009)Kite, Manga, and Gaidos]{Kite2009}
Edwin~S Kite, Michael Manga, and Eric Gaidos.
\newblock Geodynamics and rate of volcanism on massive earth-like planets.
\newblock \emph{The Astrophysical Journal}, 700\penalty0 (2):\penalty0 1732,
  2009.

\bibitem[{Kombayashi}(1967)]{Kombayashi1967}
M.~{Kombayashi}.
\newblock {Discrete equilibrium temperatures of a hypothetical planet with the
  atmosphere and the hydrosphere of one component-two phase system under
  constant solar radiation.}
\newblock \emph{J. Meteor. Soc. Japan}, 45:\penalty0 137--138, 1967.

\bibitem[{Kopparapu} et~al.(2013){Kopparapu}, {Ramirez}, {Kasting}, {Eymet},
  {Robinson}, {Mahadevan}, {Terrien}, {Domagal-Goldman}, {Meadows}, and
  {Deshpande}]{Kopparapu2013}
R.~K. {Kopparapu}, R.~{Ramirez}, J.~F. {Kasting}, V.~{Eymet}, T.~D. {Robinson},
  S.~{Mahadevan}, R.~C. {Terrien}, S.~{Domagal-Goldman}, V.~{Meadows}, and
  R.~{Deshpande}.
\newblock {Habitable Zones around Main-sequence Stars: New Estimates}.
\newblock \emph{The Astrophysical Journal}, 765:\penalty0 131, March 2013.
\newblock \doi{10.1088/0004-637X/765/2/131}.

\bibitem[{Korenaga}(2010)]{Korenaga2010}
J.~{Korenaga}.
\newblock {On the Likelihood of Plate Tectonics on Super-Earths: Does Size
  Matter?}
\newblock \emph{The Astrophysical Journal Letters}, 725:\penalty0 L43--L46,
  December 2010.
\newblock \doi{10.1088/2041-8205/725/1/L43}.

\bibitem[{Kreidberg} et~al.(2013){Kreidberg}, {Bean}, {D{\'e}sert}, {Seager},
  {Deming}, {Benneke}, {Berta}, {Stevenson}, and {Homeier}]{Kreidberg2013}
L.~{Kreidberg}, J.~{Bean}, J.~{D{\'e}sert}, S.~{Seager}, D.~{Deming},
  B.~{Benneke}, Z.~K. {Berta}, K.~B. {Stevenson}, and D.~{Homeier}.
\newblock {Transmission Spectroscopy of the Super-Earth GJ 1214b Using HST/WFC3
  in Spatial Scan Mode}.
\newblock In \emph{American Astronomical Society Meeting Abstracts}, volume 221
  of \emph{American Astronomical Society Meeting Abstracts}, page 224.03,
  January 2013.

\bibitem[{Kulikov} et~al.(2006){Kulikov}, {Lammer}, {Lichtenegger}, {Terada},
  {Ribas}, {Kolb}, {Langmayr}, {Lundin}, {Guinan}, {Barabash}, and
  {Biernat}]{Kulikov2006}
Y.~N. {Kulikov}, H.~{Lammer}, H.~I.~M. {Lichtenegger}, N.~{Terada}, I.~{Ribas},
  C.~{Kolb}, D.~{Langmayr}, R.~{Lundin}, E.~F. {Guinan}, S.~{Barabash}, and
  H.~K. {Biernat}.
\newblock {Atmospheric and water loss from early Venus}.
\newblock \emph{Planetary and Space Science}, 54:\penalty0 1425--1444, November
  2006.
\newblock \doi{10.1016/j.pss.2006.04.021}.

\bibitem[{Lammer} et~al.(2007){Lammer}, {Lichtenegger}, {Kulikov},
  {Grie{\ss}meier}, {Terada}, {Erkaev}, {Biernat}, {Khodachenko}, {Ribas},
  {Penz}, and {Selsis}]{Lammer2007}
H.~{Lammer}, H.~I.~M. {Lichtenegger}, Y.~N. {Kulikov}, {J.-M.}
  {Grie{\ss}meier}, N.~{Terada}, N.~V. {Erkaev}, H.~K. {Biernat}, M.~L.
  {Khodachenko}, I.~{Ribas}, T.~{Penz}, and F.~{Selsis}.
\newblock {Coronal Mass Ejection (CME) Activity of Low Mass M Stars as An
  Important Factor for The Habitability of Terrestrial Exoplanets. II.
  CME-Induced Ion Pick Up of Earth-like Exoplanets in Close-In Habitable
  Zones}.
\newblock \emph{Astrobiology}, 7:\penalty0 185--207, February 2007.
\newblock \doi{10.1089/ast.2006.0128}.

\bibitem[Lammer et~al.(2009)Lammer, Bredeh{\"o}ft, Coustenis, Khodachenko,
  Kaltenegger, Grasset, Prieur, Raulin, Ehrenfreund, Yamauchi,
  et~al.]{Lammer2009}
Helmut Lammer, JH~Bredeh{\"o}ft, A~Coustenis, ML~Khodachenko, L~Kaltenegger,
  O~Grasset, D~Prieur, F~Raulin, P~Ehrenfreund, M~Yamauchi, et~al.
\newblock What makes a planet habitable?
\newblock \emph{The Astronomy and Astrophysics Review}, 17\penalty0
  (2):\penalty0 181--249, 2009.

\bibitem[{Le Hir} et~al.(2008){Le Hir}, {Godd{\'e}ris}, {Donnadieu}, and
  {Ramstein}]{LeHir2008}
G.~{Le Hir}, Y.~{Godd{\'e}ris}, Y.~{Donnadieu}, and G.~{Ramstein}.
\newblock {A geochemical modelling study of the evolution of the chemical
  composition of seawater linked to a ''snowball'' glaciation}.
\newblock \emph{Biogeosciences}, 5:\penalty0 253--267, February 2008.

\bibitem[{Leconte} et~al.(2013){Leconte}, Forget, Charnay, Wordsworth, Selsis,
  and Millour]{Leconte2013}
J.~{Leconte}, F.~Forget, B.~Charnay, R.~Wordsworth, F.~Selsis, and E.~Millour.
\newblock {3D climate modeling of close-in land planets: Circulation patterns,
  climate bistability and habitability}.
\newblock \emph{submitted to Astronomy and Astrophysics}, 2013.

\bibitem[L{\'e}ger et~al.(2004)L{\'e}ger, Selsis, Sotin, Guillot, Despois,
  Mawet, Ollivier, Lab{\`e}que, Valette, Brachet, et~al.]{Leger2004}
Alain L{\'e}ger, F~Selsis, Ch~Sotin, T~Guillot, D~Despois, D~Mawet, M~Ollivier,
  A~Lab{\`e}que, C~Valette, F~Brachet, et~al.
\newblock A new family of planets?"ocean-planets".
\newblock \emph{Icarus}, 169\penalty0 (2):\penalty0 499--504, 2004.

\bibitem[{Lenardic} and {Crowley}(2012)]{Lenardic2012}
A.~{Lenardic} and J.~W. {Crowley}.
\newblock {On the Notion of Well-defined Tectonic Regimes for Terrestrial
  Planets in this Solar System and Others}.
\newblock \emph{The Astrophysical Journal}, 755:\penalty0 132, August 2012.
\newblock \doi{10.1088/0004-637X/755/2/132}.

\bibitem[Levi et~al.(2013)Levi, Sasselov, and Podolak]{Levi2013}
Amit Levi, Dimitar Sasselov, and Morris Podolak.
\newblock Volatile transport inside super-earths by entrapment in the water-ice
  matrix.
\newblock \emph{The Astrophysical Journal}, 769\penalty0 (1):\penalty0 29,
  2013.

\bibitem[Lichtenegger et~al.(2010)Lichtenegger, Lammer, Grie{\ss}meier,
  Kulikov, von Paris, Hausleitner, Krauss, and Rauer]{Lichtenegger2010}
HIM Lichtenegger, Helmut Lammer, J-M Grie{\ss}meier, Yu~N Kulikov, Philip von
  Paris, W~Hausleitner, S~Krauss, and H~Rauer.
\newblock Aeronomical evidence for higher co2 levels during earth's hadean
  epoch.
\newblock \emph{Icarus}, 210\penalty0 (1):\penalty0 1--7, 2010.

\bibitem[Lide(2000)]{CRC2000}
David~P. Lide, editor.
\newblock \emph{CRC Handbook of Chemistry and Physics}.
\newblock CRC PRESS, 81 edition, 2000.

\bibitem[Linsky et~al.(2013)Linsky, France, and Ayres]{Linsky2013}
Jeffrey~L Linsky, Kevin France, and Tom Ayres.
\newblock Computing intrinsic ly$\alpha$ fluxes of f5 v to m5 v stars.
\newblock \emph{The Astrophysical Journal}, 766\penalty0 (2):\penalty0 69,
  2013.

\bibitem[L{\'o}pez-Puertas and Taylor(2001)]{Lopez2001}
Manuel L{\'o}pez-Puertas and Fredric~William Taylor.
\newblock \emph{Non-LTE radiative transfer in the atmosphere}, volume~3.
\newblock World Scientific Publishing Company, 2001.

\bibitem[{Marcq}(2012)]{Marcq2012}
E.~{Marcq}.
\newblock {A simple 1-D radiative-convective atmospheric model designed for
  integration into coupled models of magma ocean planets}.
\newblock \emph{Journal of Geophysical Research (Planets)}, 117:\penalty0
  E01001, January 2012.
\newblock \doi{10.1029/2011JE003912}.

\bibitem[Marrero and Mason(1972)]{Marrero1972}
TR~Marrero and Edward~Allen Mason.
\newblock \emph{Gaseous diffusion coefficients}.
\newblock America Chemical Society and the American Institute of Physics, 1972.

\bibitem[{Morschhauser} et~al.(2011){Morschhauser}, {Grott}, and
  {Breuer}]{Morschhauser2011}
A.~{Morschhauser}, M.~{Grott}, and D.~{Breuer}.
\newblock {Crustal recycling, mantle dehydration, and the thermal evolution of
  Mars}.
\newblock \emph{Icarus}, 212:\penalty0 541--558, April 2011.
\newblock \doi{10.1016/j.icarus.2010.12.028}.

\bibitem[Mota et~al.(2005)Mota, Parafita, Giuliani, Hubin-Franskin, Lourenco,
  Garcia, Hoffmann, Mason, Ribeiro, Raposo, et~al.]{Mota2005}
R~Mota, R~Parafita, A~Giuliani, M-J Hubin-Franskin, JMC Lourenco, G~Garcia,
  SV~Hoffmann, NJ~Mason, PA~Ribeiro, M~Raposo, et~al.
\newblock Water vuv electronic state spectroscopy by synchrotron radiation.
\newblock \emph{Chemical physics letters}, 416\penalty0 (1):\penalty0 152--159,
  2005.

\bibitem[Murray-Clay et~al.(2009)Murray-Clay, Chiang, and Murray]{Murray2009}
Ruth~A Murray-Clay, Eugene~I Chiang, and Norman Murray.
\newblock Atmospheric escape from hot jupiters.
\newblock \emph{The Astrophysical Journal}, 693\penalty0 (1):\penalty0 23,
  2009.

\bibitem[{O'Brien} et~al.(2006){O'Brien}, {Morbidelli}, and
  {Levison}]{OBrien2006}
D.~P. {O'Brien}, A.~{Morbidelli}, and H.~F. {Levison}.
\newblock {Terrestrial planet formation with strong dynamical friction}.
\newblock \emph{Icarus}, 184:\penalty0 39--58, September 2006.
\newblock \doi{10.1016/j.icarus.2006.04.005}.

\bibitem[{O'Neill} and {Lenardic}(2007)]{ONeill2007}
C.~{O'Neill} and A.~{Lenardic}.
\newblock {Geological consequences of super-sized Earths}.
\newblock \emph{Geophysical Research Letters}, 34:\penalty0 L19204, October
  2007.
\newblock \doi{10.1029/2007GL030598}.

\bibitem[{O'Rourke} and {Korenaga}(2012)]{ORourke2012}
J.~G. {O'Rourke} and J.~{Korenaga}.
\newblock {Terrestrial planet evolution in the stagnant-lid regime: Size
  effects and the formation of self-destabilizing crust}.
\newblock \emph{Icarus}, 221:\penalty0 1043--1060, November 2012.
\newblock \doi{10.1016/j.icarus.2012.10.015}.

\bibitem[Pierrehumbert(2011)]{Pierrehumbert2011b}
Raymond~T Pierrehumbert.
\newblock {A palette of climates for Gliese 581g}.
\newblock \emph{The Astrophysical Journal Letters}, 726\penalty0 (1):\penalty0
  L8, 2011.

\bibitem[Pierrehumbert et~al.(2007)Pierrehumbert, Brogniez, Roca,
  et~al.]{Pierrehumbert2007}
Raymond~T Pierrehumbert, H{\'e}l{\`e}ne Brogniez, R{\'e}my Roca, et~al.
\newblock On the relative humidity of the earth's atmosphere, 2007.

\bibitem[Pierrehumbert(2010)]{Pierrehumbert2011BOOK}
R.T. Pierrehumbert.
\newblock \emph{Principles of Planetary Climate}.
\newblock Cambridge University Press, 2010.
\newblock ISBN 9780521865562.
\newblock URL \url{http://books.google.com/books?id=bO\_U8f5pVR8C}.

\bibitem[Pope et~al.(2012)Pope, Bird, and Rosing]{Pope2012}
Emily~C. Pope, Dennis~K. Bird, and Minik~T. Rosing.
\newblock Isotope composition and volume of earth's early oceans.
\newblock \emph{Proceedings of the National Academy of Sciences}, 2012.
\newblock \doi{10.1073/pnas.1115705109}.

\bibitem[{Raymond} et~al.(2006){Raymond}, {Quinn}, and {Lunine}]{Raymond2006}
S.~N. {Raymond}, T.~{Quinn}, and J.~I. {Lunine}.
\newblock {High-resolution simulations of the final assembly of Earth-like
  planets I. Terrestrial accretion and dynamics}.
\newblock \emph{Icarus}, 183:\penalty0 265--282, August 2006.
\newblock \doi{10.1016/j.icarus.2006.03.011}.

\bibitem[{Ribas} et~al.(2005){Ribas}, {Guinan}, {G{\"u}del}, and
  {Audard}]{Ribas2005}
I.~{Ribas}, E.~F. {Guinan}, M.~{G{\"u}del}, and M.~{Audard}.
\newblock {Evolution of the Solar Activity over Time and Effects on Planetary
  Atmospheres. I. High-Energy Irradiances (1-1700 {\AA})}.
\newblock \emph{The Astrophysical Journal}, 622:\penalty0 680--694, 2005.

\bibitem[Ribas et~al.(2010)Ribas, de~Mello, Ferreira, H{\'e}brard, Selsis,
  Catal{\'a}n, Garc{\'e}s, do~Nascimento~Jr, and De~Medeiros]{Ribas2010}
I~Ribas, GF~Porto de~Mello, LD~Ferreira, E~H{\'e}brard, Franck Selsis,
  S~Catal{\'a}n, A~Garc{\'e}s, JD~do~Nascimento~Jr, and JR~De~Medeiros.
\newblock {Evolution of the Solar Activity Over Time and Effects on Planetary
  Atmospheres. II. $\kappa$1 Ceti, an Analog of the Sun when Life Arose on
  Earth}.
\newblock \emph{The Astrophysical Journal}, 714\penalty0 (1):\penalty0 384,
  2010.

\bibitem[{Rothman} et~al.(2009){Rothman}, {Gordon}, {Barbe}, {Benner},
  {Bernath}, {Birk}, {Boudon}, {Brown}, {Campargue}, {Champion}, {Chance},
  {Coudert}, {Dana}, {Devi}, {Fally}, {Flaud}, {Gamache}, {Goldman},
  {Jacquemart}, {Kleiner}, {Lacome}, {Lafferty}, {Mandin}, {Massie},
  {Mikhailenko}, {Miller}, {Moazzen-Ahmadi}, {Naumenko}, {Nikitin}, {Orphal},
  {Perevalov}, {Perrin}, {Predoi-Cross}, {Rinsland}, {Rotger}, {{\v S}ime{\v
  c}kov{\'a}}, {Smith}, {Sung}, {Tashkun}, {Tennyson}, {Toth}, {Vandaele}, and
  {Vander Auwera}]{Rothman2009}
L.~S. {Rothman}, I.~E. {Gordon}, A.~{Barbe}, D.~C. {Benner}, P.~F. {Bernath},
  M.~{Birk}, V.~{Boudon}, L.~R. {Brown}, A.~{Campargue}, J.-P. {Champion},
  K.~{Chance}, L.~H. {Coudert}, V.~{Dana}, V.~M. {Devi}, S.~{Fally}, J.-M.
  {Flaud}, R.~R. {Gamache}, A.~{Goldman}, D.~{Jacquemart}, I.~{Kleiner},
  N.~{Lacome}, W.~J. {Lafferty}, J.-Y. {Mandin}, S.~T. {Massie}, S.~N.
  {Mikhailenko}, C.~E. {Miller}, N.~{Moazzen-Ahmadi}, O.~V. {Naumenko}, A.~V.
  {Nikitin}, J.~{Orphal}, V.~I. {Perevalov}, A.~{Perrin}, A.~{Predoi-Cross},
  C.~P. {Rinsland}, M.~{Rotger}, M.~{{\v S}ime{\v c}kov{\'a}}, M.~A.~H.
  {Smith}, K.~{Sung}, S.~A. {Tashkun}, J.~{Tennyson}, R.~A. {Toth}, A.~C.
  {Vandaele}, and J.~{Vander Auwera}.
\newblock {The HITRAN 2008 molecular spectroscopic database}.
\newblock \emph{Journal of Quantitative Spectroscopy and Radiative Transfer},
  110:\penalty0 533--572, 2009.
\newblock \doi{10.1016/j.jqsrt.2009.02.013}.

\bibitem[Russell et~al.(1994)Russell, Daniel, Hall, and
  Sherringham]{Russell1994}
Michael~J Russell, Roy~M Daniel, Allan~J Hall, and John~A Sherringham.
\newblock A hydrothermally precipitated catalytic iron sulphide membrane as a
  first step toward life.
\newblock \emph{Journal of Molecular Evolution}, 39\penalty0 (3):\penalty0
  231--243, 1994.

\bibitem[{Segura} et~al.(2003){Segura}, {Krelove}, {Kasting}, {Sommerlatt},
  {Meadows}, {Crisp}, {Cohen}, and {Mlawer}]{Segura2003}
A.~{Segura}, K.~{Krelove}, J.~F. {Kasting}, D.~{Sommerlatt}, V.~{Meadows},
  D.~{Crisp}, M.~{Cohen}, and E.~{Mlawer}.
\newblock {Ozone Concentrations and Ultraviolet Fluxes on Earth-Like Planets
  Around Other Stars}.
\newblock \emph{Astrobiology}, 3:\penalty0 689--708, December 2003.
\newblock \doi{10.1089/153110703322736024}.

\bibitem[{Segura} et~al.(2002){Segura}, {Toon}, {Colaprete}, and
  {Zahnle}]{Segura2002}
T.~L. {Segura}, O.~B. {Toon}, A.~{Colaprete}, and K.~{Zahnle}.
\newblock {Environmental Effects of Large Impacts on Mars}.
\newblock \emph{Science}, 298:\penalty0 1977--1980, December 2002.

\bibitem[Segura et~al.(2008)Segura, Toon, and Colaprete]{Segura2008}
Teresa~L Segura, O~Brian Toon, and Anthony Colaprete.
\newblock {Modeling the environmental effects of moderate-sized impacts on
  Mars}.
\newblock \emph{Journal of Geophysical Research: Planets (1991--2012)},
  113\penalty0 (E11), 2008.

\bibitem[Selsis et~al.(2002)Selsis, Despois, and Parisot]{Selsis2002}
F~Selsis, D~Despois, and J-P Parisot.
\newblock {Signature of life on exoplanets: Can Darwin produce false positive
  detections?}
\newblock \emph{Astronomy and Astrophysics}, 388\penalty0 (3):\penalty0
  985--1003, 2002.

\bibitem[{Selsis} et~al.(2007){Selsis}, {Kasting}, {Levrard}, {Paillet},
  {Ribas}, and {Delfosse}]{Selsis2007}
F.~{Selsis}, J.~F. {Kasting}, B.~{Levrard}, J.~{Paillet}, I.~{Ribas}, and
  X.~{Delfosse}.
\newblock {Habitable planets around the star Gliese 581?}
\newblock \emph{Astron. Astrophys.}, 476:\penalty0 1373--1387, December 2007.

\bibitem[Selsis et~al.(2011)Selsis, Wordsworth, and Forget]{Selsis2011}
F~Selsis, RD~Wordsworth, and F~Forget.
\newblock Thermal phase curves of nontransiting terrestrial exoplanets: I.
  characterizing atmospheres.
\newblock \emph{Astronomy \& Astrophysics}, 532, 2011.

\bibitem[Shine et~al.(2012)Shine, Ptashnik, and R{\"a}del]{Shine2012}
Keith~P Shine, Igor~V Ptashnik, and Gaby R{\"a}del.
\newblock The water vapour continuum: brief history and recent developments.
\newblock \emph{Surveys in geophysics}, 33\penalty0 (3-4):\penalty0 535--555,
  2012.

\bibitem[{Sleep} and {Zahnle}(2001)]{Sleep2001}
N.~H. {Sleep} and K.~{Zahnle}.
\newblock {Carbon dioxide cycling and implications for climate on ancient
  Earth}.
\newblock \emph{Journal of Geophysical Research}, 106:\penalty0 1373--1400,
  January 2001.
\newblock \doi{10.1029/2000JE001247}.

\bibitem[Stark et~al.(1992)Stark, Smith, Huber, Yoshino, Stevens, and
  Ito]{Stark1992}
G~Stark, Peter~L Smith, KP~Huber, K~Yoshino, MH~Stevens, and K~Ito.
\newblock Absorption band oscillator strengths of n transitions between 95.8
  and 99.4 nm.
\newblock \emph{The Journal of chemical physics}, 97:\penalty0 4809, 1992.

\bibitem[Tarduno et~al.(2010)Tarduno, Cottrell, Watkeys, Hofmann, Doubrovine,
  Mamajek, Liu, Sibeck, Neukirch, and Usui]{Tarduno2010}
John~A Tarduno, Rory~D Cottrell, Michael~K Watkeys, Axel Hofmann, Pavel~V
  Doubrovine, Eric~E Mamajek, Dunji Liu, David~G Sibeck, Levi~P Neukirch, and
  Yoichi Usui.
\newblock Geodynamo, solar wind, and magnetopause 3.4 to 3.45 billion years
  ago.
\newblock \emph{Science}, 327\penalty0 (5970):\penalty0 1238--1240, 2010.

\bibitem[Thuillier et~al.(2004)Thuillier, Floyd, Woods, Cebula, Hilsenrath,
  Hers{\'e}, et~al.]{Thuillier2004}
G{\'e}rard Thuillier, Linton Floyd, Thomas~N Woods, Richard Cebula, Ernest
  Hilsenrath, Michel Hers{\'e}, et~al.
\newblock Solar irradiance reference spectra.
\newblock \emph{Geophysical Monograph Series}, 141:\penalty0 171--194, 2004.

\bibitem[{Tian}(2009)]{Tian2009}
F.~{Tian}.
\newblock {Thermal Escape from Super Earth Atmospheres in the Habitable Zones
  of M Stars}.
\newblock \emph{The Astrophysical Journal}, 703:\penalty0 905--909, September
  2009.
\newblock \doi{10.1088/0004-637X/703/1/905}.

\bibitem[{Tian} et~al.(2009){Tian}, {Kasting}, and {Solomon}]{Tian2009b}
F.~{Tian}, J.~F. {Kasting}, and S.~C. {Solomon}.
\newblock {Thermal escape of carbon from the early Martian atmosphere}.
\newblock \emph{Geophysical Research Letters}, 36:\penalty0 L02205, January
  2009.
\newblock \doi{10.1029/2008GL036513}.

\bibitem[Titov et~al.(2007)Titov, Bullock, Crisp, Renno, Taylor, and
  Zasova]{Titov2007}
Dmitry~V Titov, Mark~A Bullock, David Crisp, Nilton~O Renno, Fredric~W Taylor,
  and Ljudmilla~V Zasova.
\newblock {Radiation in the atmosphere of Venus}.
\newblock \emph{GEOPHYSICAL MONOGRAPH-AMERICAN GEOPHYSICAL UNION},
  176:\penalty0 121, 2007.

\bibitem[{Toon} et~al.(1989){Toon}, {McKay}, {Ackerman}, and
  {Santhanam}]{Toon1989}
O.~B. {Toon}, C.~P. {McKay}, T.~P. {Ackerman}, and K.~{Santhanam}.
\newblock {Rapid calculation of radiative heating rates and photodissociation
  rates in inhomogeneous multiple scattering atmospheres}.
\newblock \emph{JGR}, 94:\penalty0 16287--16301, November 1989.

\bibitem[{Valencia} et~al.(2007){Valencia}, {O'Connell}, and
  {Sasselov}]{Valencia2007}
D.~{Valencia}, R.~J. {O'Connell}, and D.~D. {Sasselov}.
\newblock {Inevitability of Plate Tectonics on Super-Earths}.
\newblock \emph{The Astrophysical Journal}, 670:\penalty0 L45--L48, November
  2007.

\bibitem[{von Paris} et~al.(2010){von Paris}, {Gebauer}, {Godolt}, {Grenfell},
  {Hedelt}, {Kitzmann}, {Patzer}, {Rauer}, and {Stracke}]{vonParis2010}
P.~{von Paris}, S.~{Gebauer}, M.~{Godolt}, J.~L. {Grenfell}, P.~{Hedelt},
  D.~{Kitzmann}, A.~B.~C. {Patzer}, H.~{Rauer}, and B.~{Stracke}.
\newblock {The extrasolar planet Gliese 581d: a potentially habitable planet?}
\newblock \emph{Astronomy and Astrophysics}, 522:\penalty0 A23+, November 2010.
\newblock \doi{10.1051/0004-6361/201015329}.

\bibitem[{Walker} et~al.(1981){Walker}, {Hays}, and {Kasting}]{Walker1981}
J.~C.~G. {Walker}, P.~B. {Hays}, and J.~F. {Kasting}.
\newblock {A negative feedback mechanism for the long-term stabilization of the
  earth's surface temperature}.
\newblock \emph{Journal of Geophysical Research}, 86:\penalty0 9776--9782,
  October 1981.
\newblock \doi{10.1029/JC086iC10p09776}.

\bibitem[{West} et~al.(2005){West}, {Galy}, and {Bickle}]{West2005}
A.~J. {West}, A.~{Galy}, and M.~{Bickle}.
\newblock {Tectonic and climatic controls on silicate weathering [rapid
  communication]}.
\newblock \emph{Earth and Planetary Science Letters}, 235:\penalty0 211--228,
  June 2005.
\newblock \doi{10.1016/j.epsl.2005.03.020}.

\bibitem[{Wordsworth}(2012)]{Wordsworth2012a}
R.~{Wordsworth}.
\newblock {Transient conditions for biogenesis on low-mass exoplanets with
  escaping hydrogen atmospheres}.
\newblock \emph{Icarus}, 219:\penalty0 267--273, May 2012.
\newblock \doi{10.1016/j.icarus.2012.02.035}.

\bibitem[{Wordsworth} et~al.(2010{\natexlab{a}}){Wordsworth}, {Forget}, and
  {Eymet}]{Wordsworth2010}
R.~{Wordsworth}, F.~{Forget}, and V.~{Eymet}.
\newblock {Infrared collision-induced and far-line absorption in dense CO2
  atmospheres}.
\newblock \emph{Icarus}, 210:\penalty0 992--997, December 2010{\natexlab{a}}.
\newblock \doi{10.1016/j.icarus.2010.06.010}.

\bibitem[{Wordsworth} et~al.(2010{\natexlab{b}}){Wordsworth}, {Forget},
  {Selsis}, {Madeleine}, {Millour}, and {Eymet}]{Wordsworth2010b}
R.~D. {Wordsworth}, F.~{Forget}, F.~{Selsis}, {J.-B.} {Madeleine},
  E.~{Millour}, and V.~{Eymet}.
\newblock {Is Gliese 581d habitable? Some constraints from radiative-convective
  climate modeling}.
\newblock \emph{Astronomy and Astrophysics}, 522:\penalty0 A22+,
  2010{\natexlab{b}}.
\newblock \doi{10.1051/0004-6361/201015053}.

\bibitem[{Wordsworth} et~al.(2011){Wordsworth}, {Forget}, {Selsis}, {Millour},
  {Charnay}, and {Madeleine}]{Wordsworth2011}
R.~D. {Wordsworth}, F.~{Forget}, F.~{Selsis}, E.~{Millour}, B.~{Charnay}, and
  J.-B. {Madeleine}.
\newblock {Gliese 581d is the First Discovered Terrestrial-mass Exoplanet in
  the Habitable Zone}.
\newblock \emph{The Astrophysical Journal Letters}, 733:\penalty0 L48, June
  2011.
\newblock \doi{10.1088/2041-8205/733/2/L48}.

\bibitem[Wordsworth and Pierrehumbert(2013)]{Wordsworth2013b}
Robin Wordsworth and Raymond Pierrehumbert.
\newblock Hydrogen-nitrogen greenhouse warming in earth's early atmosphere.
\newblock \emph{Science}, 339\penalty0 (6115):\penalty0 64--67, 2013.

\bibitem[Zahnle et~al.(1988)Zahnle, Kasting, and Pollack]{Zahnle1988}
Kevin~J Zahnle, James~F Kasting, and James~B Pollack.
\newblock Evolution of a steam atmosphere during earth's accretion.
\newblock \emph{Icarus}, 74\penalty0 (1):\penalty0 62--97, 1988.

\bibitem[Zsom et~al.(2013)Zsom, Seager, and de~Wit]{Zsom2013}
Andras Zsom, Sara Seager, and Julien de~Wit.
\newblock Towards the minimum inner edge distance of the habitable zone.
\newblock \emph{arXiv preprint arXiv:1304.3714}, 2013.

\end{thebibliography}
\bibliographystyle{plainnat}

\clearpage

\begin{figure}[h]
	\begin{center}
		{\includegraphics[width=4.5in]{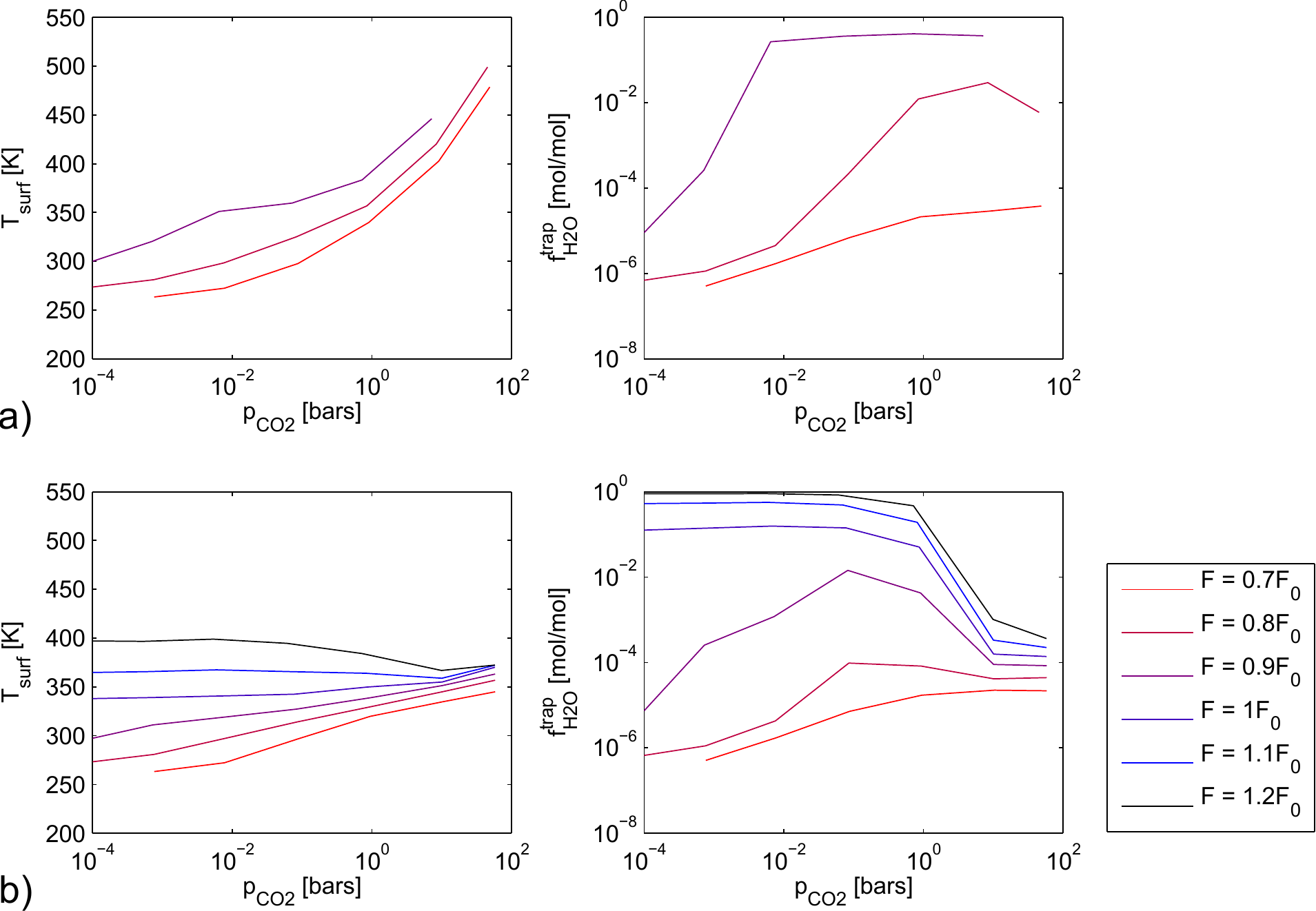}}
	\end{center}
	\caption{a-b) As for Figure~\ref{fig:vstrat}b-c), but assuming an M-star incident spectrum. In cases where no data is shown, an equilbrium solution
was not found for any surface temperature between 250 and 500~K.}
	\label{fig:vstrat2}
\end{figure}

\begin{figure}[h]
	\begin{center}
		{\includegraphics[width=3.5in]{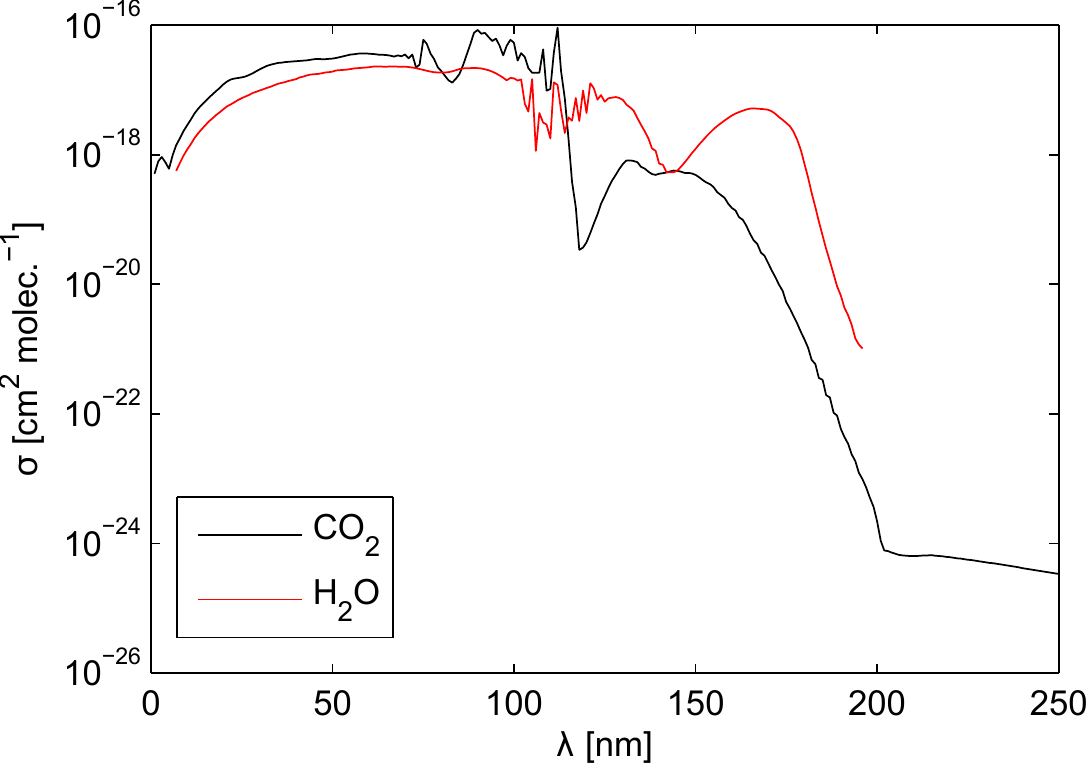}}
	\end{center}
	\caption{CO$_2$ and H$_2$O absorption cross-sections in the UV used in the model, as a function of wavelength.}
	\label{fig:CO2_H2O_XUV}
\end{figure}

\begin{figure}[h]
	\begin{center}
		{\includegraphics[width=6in]{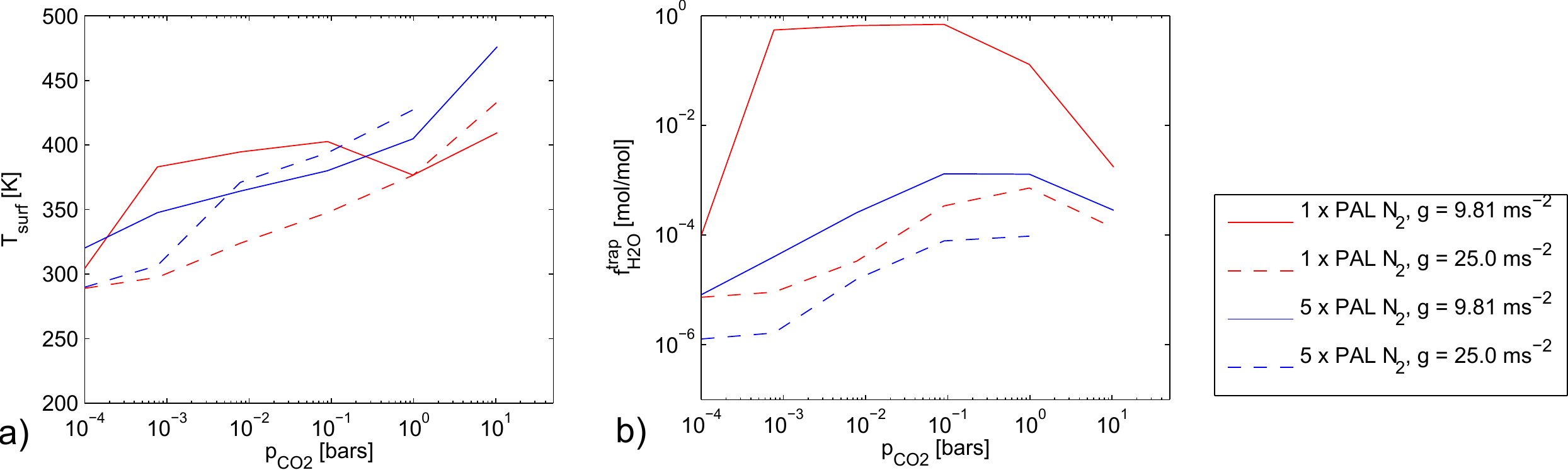}}
	\end{center}
	\caption{a) Surface temperature and b) stratospheric H$_2$O volume mixing ratio as a function of surface CO$_2$ partial pressure,
for simulations with varying surface gravity and total atmospheric nitrogen content.}
	\label{fig:varyN2g}
\end{figure}

\begin{figure}[h]
	\begin{center}
		{\includegraphics[width=3.5in]{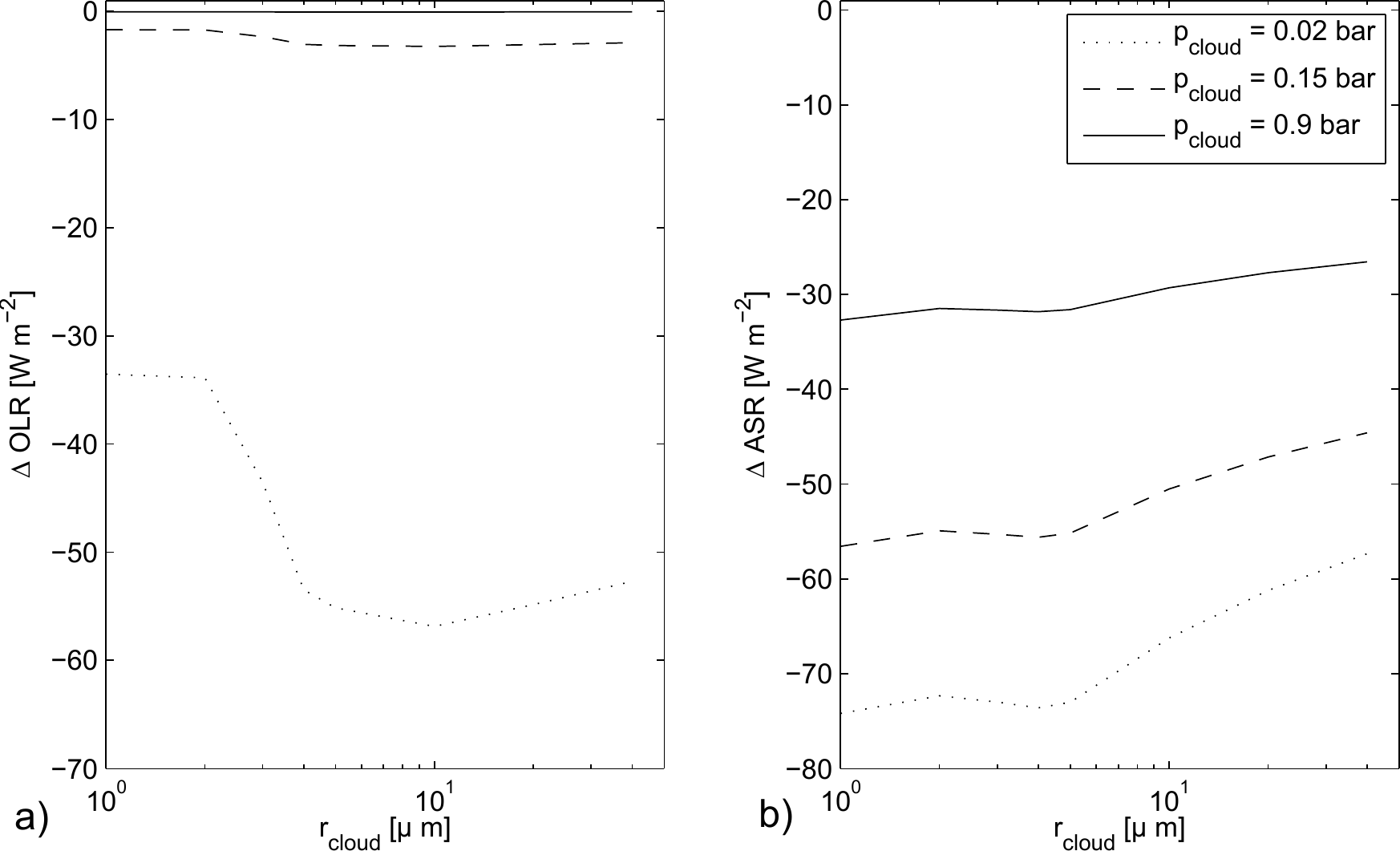}}
	\end{center}
	\caption{Radiative effects of clouds for an atmosphere with the same composition, temperature profile and stellar forcing as shown in Fig.~\ref{fig:radheat}. a) Longwave and b) shortwave radiative forcing vs. the clear-sky case as a function of H$_2$O cloud particle radius, for a single layer with 100\% coverage and opacity $\tau=1.0$ at 1.5~$\mu m$.}
	\label{fig:clouds}
\end{figure}

\begin{figure}[h]
	\begin{center}
		{\includegraphics[width=5in]{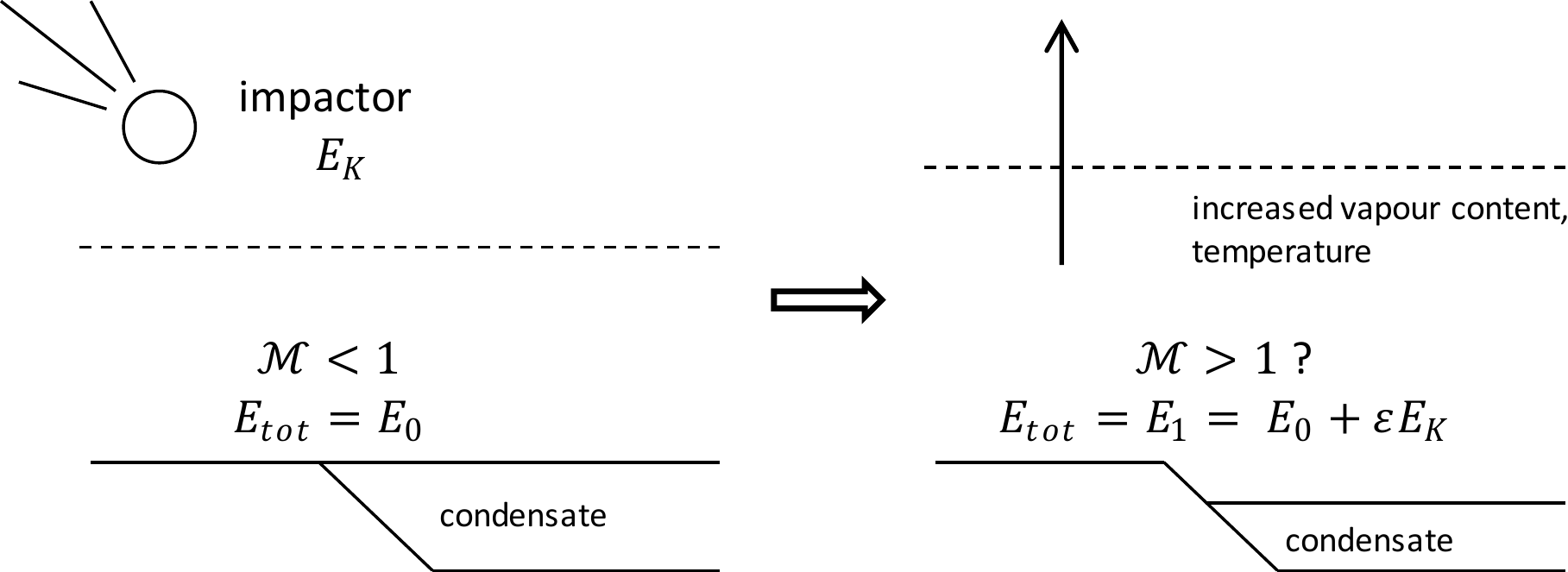}}
	\end{center}
	\caption{Schematic of the effect of an impact on a planet with a dense atmosphere. Some of the impactor kinetic energy is used to convert
surface condensate material (here, liquid water) to vapour in the atmosphere. If the impactor radius is large enough, this may heat the atmosphere enough to moisten the stratosphere and allow transitory periods of rapid H$_2$O photolysis. However, large impactors will also cause substantial amounts of the atmosphere to be ejected to space.}
	\label{fig:impactschematic}
\end{figure}

\begin{figure}[h]
	\begin{center}
		{\includegraphics[width=3.5in]{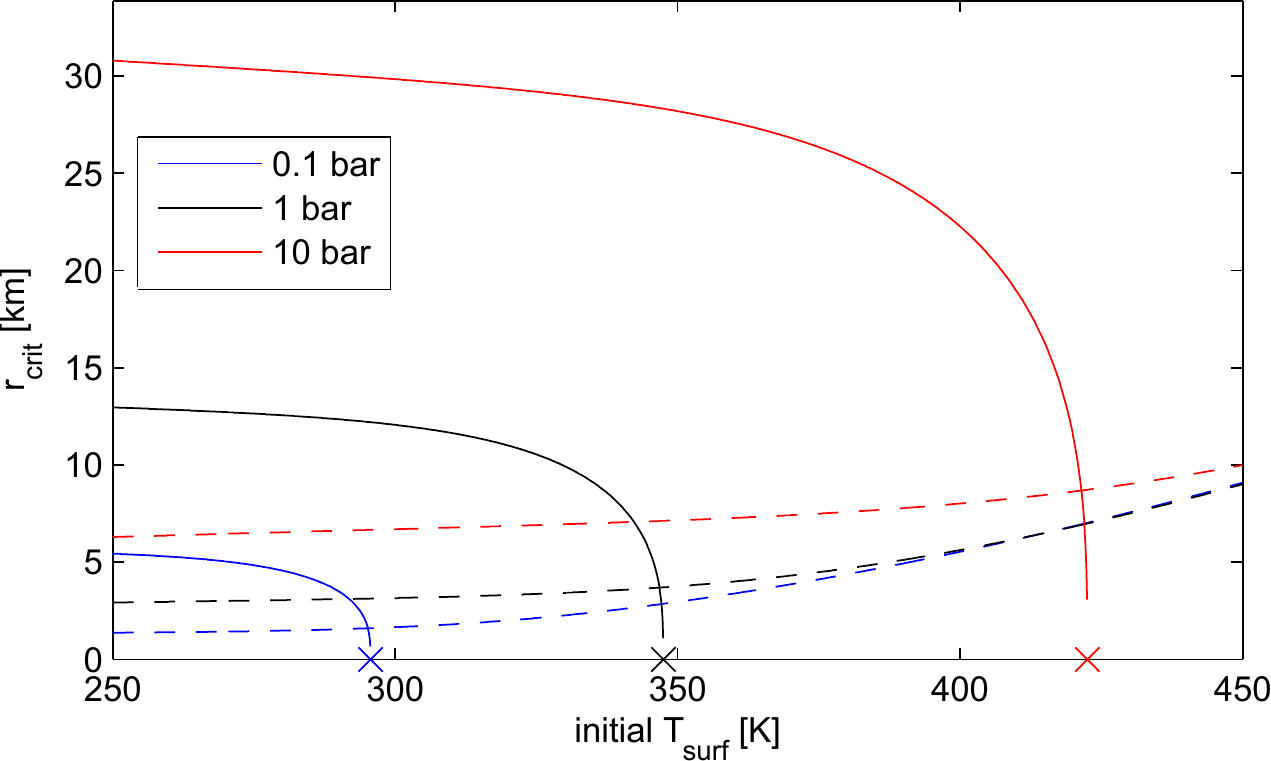}}
	\end{center}
	\caption{Critical impactor radius necessary to cause transition to a $\mathcal M=1$ moist stratosphere regime assuming 100\% energy conversion efficiency (solid lines), and to cause significant atmospheric erosion to space (dashed lines). Colors indicate the partial pressure of the incondensible gas (assumed 100\% CO$_2$ here for simplicity). Crosses at the base of the plot indicate the temperature $T_{surf}^*$ at which $\mathcal M=1$. The increase of the critical erosion radius with $T_{surf}$ is due to its dependence on the scale height of the atmosphere.}
	\label{fig:impactimpact}
\end{figure}

\begin{figure}[h]
	\begin{center}
		{\includegraphics[width=3.5in]{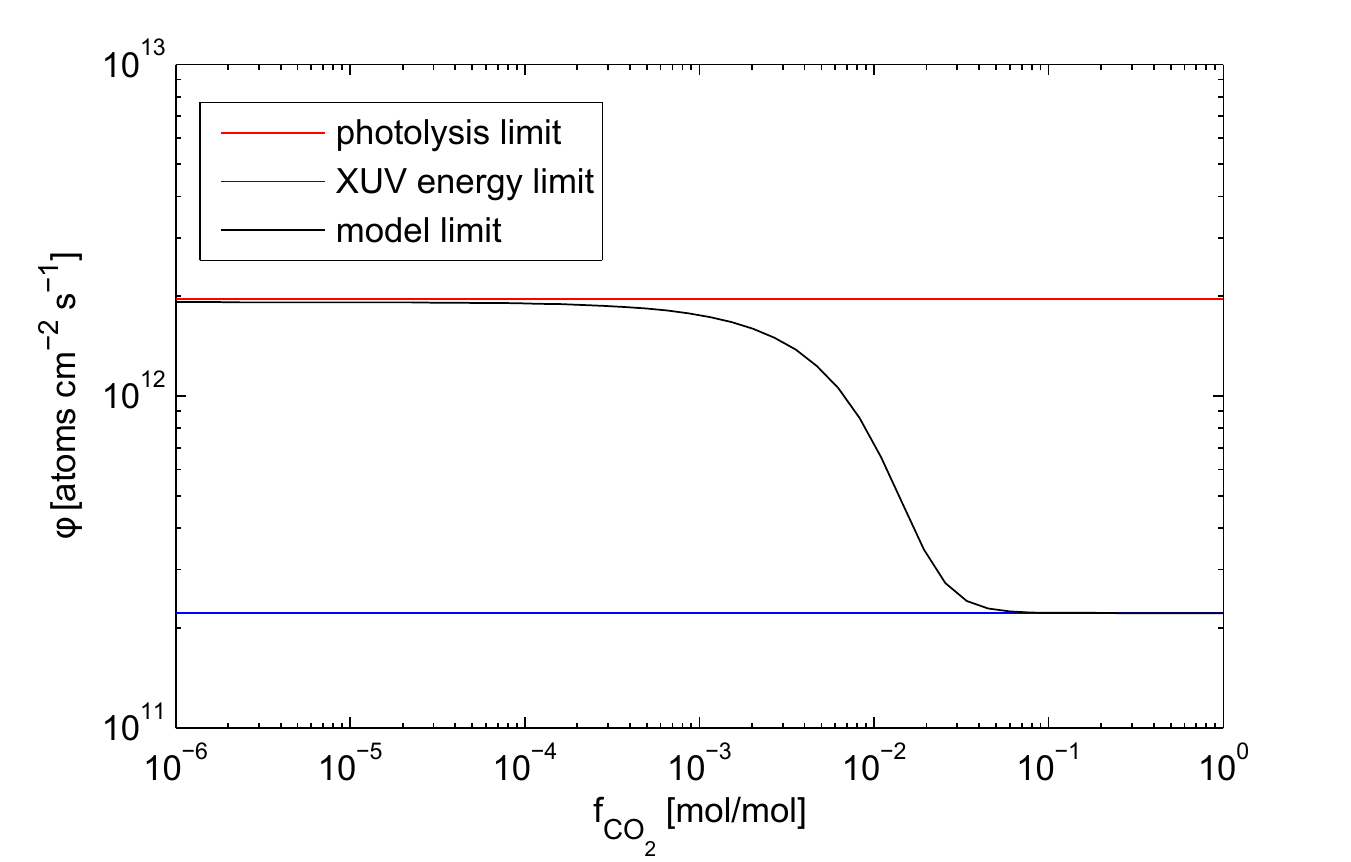}}
	\end{center}
	\caption{Hydrogen (H) escape rate as a function of CO$_2$ volume mixing ratio (molar concentration) for a CO$_2$/H$_2$O atmosphere
under G-class stellar insolation with Sun-like XUV/UV spectrum.}
	\label{fig:CO2_H2O_escapebox}
\end{figure}

\begin{figure}[h]
	\begin{center}
		{\includegraphics[width=5in]{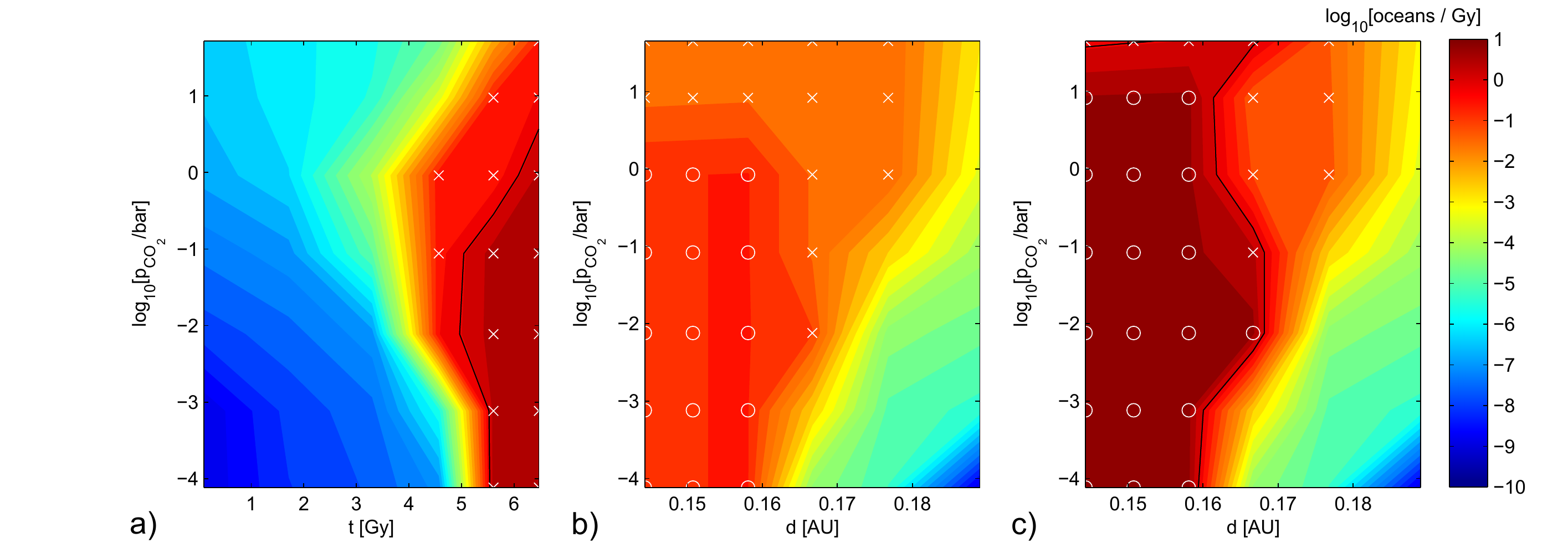}}
	\end{center}
	\caption{a) Water loss rate as a function of surface CO$_2$ partial pressure and time for an Earth-like planet around a G-star at 1~AU. Water loss rate as a function of surface CO$_2$ partial pressure and orbital distance for an Earth-like planet around b) a moderately active M3 class star (GJ~436) and c) a star with elevated Lyman-$\alpha$ emission. {White crosses / circles indicate data points where escape was energy / photolysis rate limited, respectively. The solid black line indicates the contour for a loss rate of 1~Earth~ocean~Gy$^{-1}$}.}
	\label{fig:escape_total}
\end{figure}

\begin{figure}[h]
	\begin{center}
		{\includegraphics[width=3.5in]{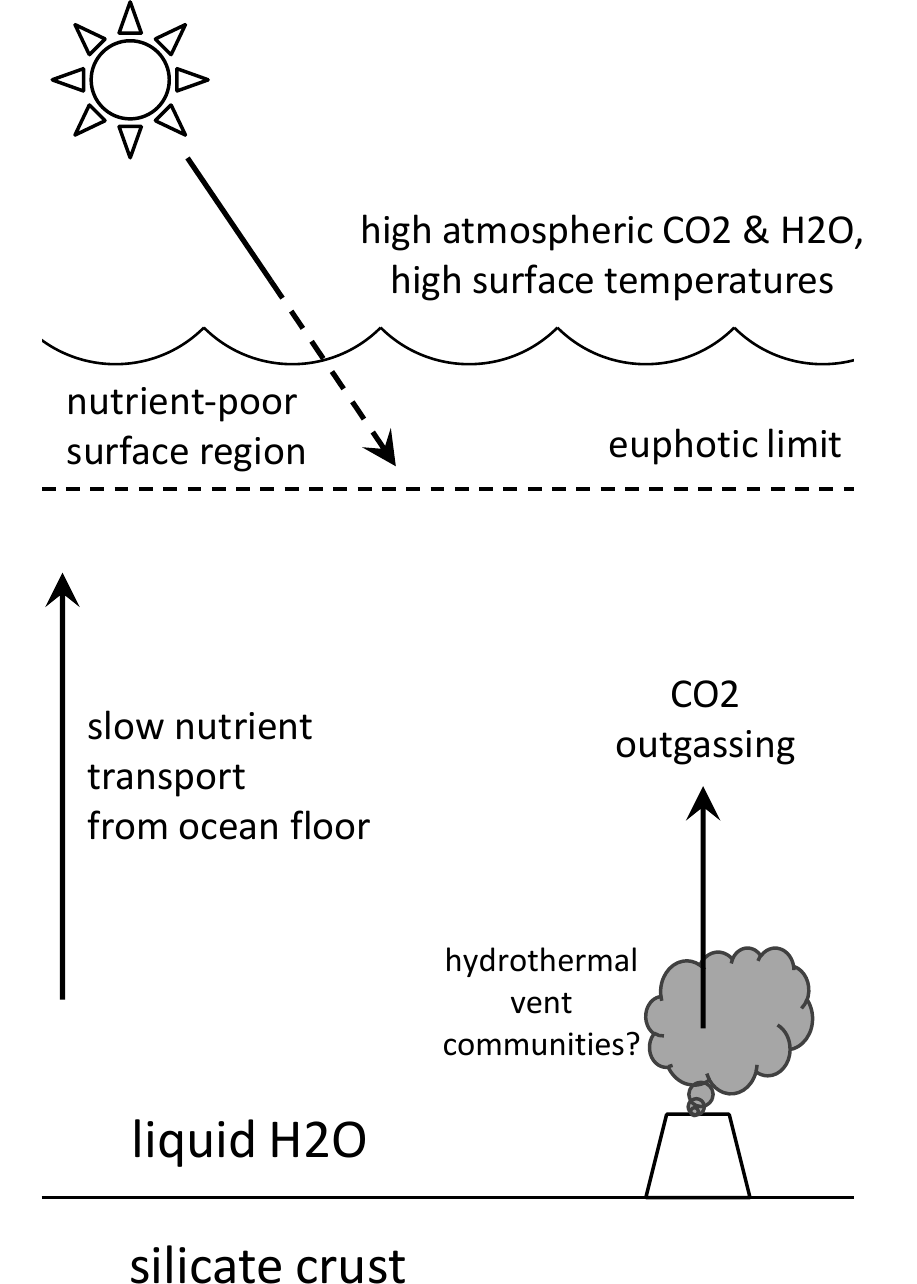}}
	\end{center}
	\caption{Schematic of processes affecting climate and biospheric productivity on a hypothetical Earth-like planet with oceans deep enough to cover the entire surface.}
	\label{fig:oceanschematic}
\end{figure}

\begin{figure}[h]
	\begin{center}
		{\includegraphics[width=3.5in]{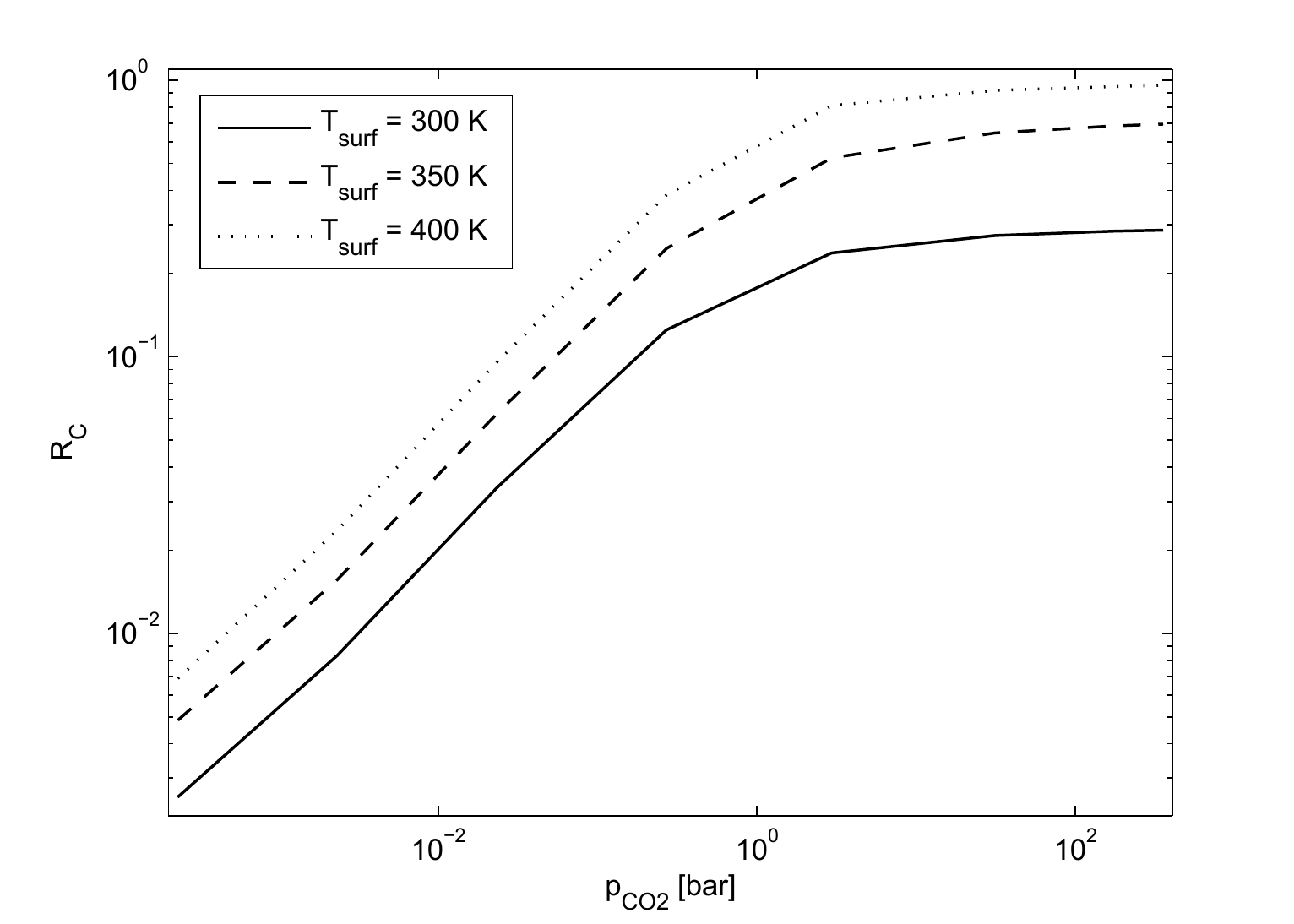}}
	\end{center}
	\caption{Ratio between total atmosphere and ocean inorganic carbon as a function of surface temperature and CO$_2$ partial pressure, for a predominately rocky super-Earth with 10 times Earth's surface liquid water content, surface gravity $15$~m~s$^{-2}$ and radius $1.3 r_E$.}
	\label{fig:ocean_C_content}
\end{figure}

\end{document}